\documentclass[preprintnumbers,secnumarabic,amsmath,amssymb,nofootinbib,floatfix]{revtex4}
\usepackage{graphics}
\usepackage{graphicx}% Include figure files
\usepackage{dcolumn}% Align table columns on decimal point
\usepackage{bm}% bold math
\usepackage{mathrsfs}
\usepackage{slashed}
\usepackage{amsmath}
\usepackage{epsfig}
\usepackage{amsfonts}
\usepackage{amssymb}
\usepackage{tikz}
\usepackage{tikz-cd}

\newcommand{\beq}{\begin{equation}}
\newcommand{\eeq}{\end{equation}}
\newcommand{\bea}{\begin{eqnarray}}
\newcommand{\eea}{\end{eqnarray}}
\newcommand{\nn}{\nonumber}
\begin{document}

\title{ Leading singularities in higher-derivative Yang-Mills theory and quadratic gravity }
\author{Gabriel Menezes}
\email{gabrielmenezes@ufrrj.br}
\affiliation{~\\ Departamento de F\'{i}sica, Universidade Federal Rural do Rio de Janeiro, 23897-000, Serop\'{e}dica, RJ, Brazil \\ }
%

%%%%%%%%%%%%%%%%%%%%%%%%%
\begin{abstract}
In this work we explore general leading singularities of one-loop amplitudes in higher-derivative Yang-Mills and quadratic gravity. These theories are known to possess propagators which contain quadratic and quartic momentum dependence, which leads to the presence of an unstable ghostlike resonance. However, unitarity cuts are not to be taken through unstable particles and therefore unitarity is still satisfied. On the other hand, this could engender issues when calculating leading singularities which are generalizations of unitarity cuts. Nevertheless, we will show with explicit examples how 
leading singularities are still well defined and accordingly they are able to capture relevant information on the analytic structure of amplitudes in such higher-derivative theories. We discuss some simple one-loop amplitudes which clarify these features.
\end{abstract}
%%%%%%%%%%%%%%%%%%%%%%%%%

%\newpage
\maketitle

%%%%%%%%%%%%%%%%%%%%%%%%%
\section{Introduction}

Analytic properties of scattering amplitudes have been meticulously examined in the literature~\cite{Witten:2003nn,Britto:2004nc,Arkani-Hamed:2008owk,Arkani-Hamed:2009kmp,Arkani-Hamed:2010zjl,book,Eden:1966dnq,Weinberg:1964ev,Weinberg:1964ew,Weinberg:1965rz,Olive:65,book2}. The idea is to regard the amplitude as an analytic function of kinematical invariants -- and physical input would be needed to ascertain all associated singularities. Several decades ago, this project developed into the ambitious expectation that an adequate understanding of the structure of its singularities would enable one to determine the S-matrix itself and, as a consequence, one would obtain a deep insight of all the interactions involved. However, this program has proven to be a formidable task. In any case, such investigations have unveiled the intricate analytic structure amplitudes may have in perturbation theory. Indeed, at loop level Feynman diagrams enjoy a configuration rather involved consisting of nested branch cuts. This complicated structure gets even more complex when one goes to higher orders in perturbation theory. This non-trivial analytic structure can be envisaged when assuming unitarity; that is, imposing the generalized optical theorem should be satisfied. It is the assumption of this constraint that makes it possible to relate  the discontinuity of an amplitude to the exchange of on-shell states amongst sets of external particles. One-particle exchange is associated with the existence of poles whereas a two-particle exchange indicates the presence of a branch cut. 

The discontinuity across a pole is the residue of the associated holomorphic function at such a pole. Often we discuss discontinuities in a given channel from the perspective of the so-called cutting rules~\cite{Cutkosky:1960sp,Schwartz:13}, which basically deals with two-particle exchanges. In summary, we use such rules to compute the imaginary part of loop amplitudes, which furnishes their discontinuities. An alternative interpretation is to conceive unitarity cuts as residues of the amplitude; for instance, for two-particle exchange we take two propagators $1/(L_1^2 - m_1^2)$ and $1/(L_2^2 - m_2^2)$ (they can also be massless), then integrate over contours that surround the poles $L_i^2 -m_i^2 = 0$ in the associated complex planes. This amounts to removing the principal part of the propagator, keeping only its imaginary part given by the associated delta functions $\delta(L_i^2 -m_i^2)$, in compliance with the fact that imaginary parts of loop amplitudes correspond to intermediate particles going on-shell. In this regard, the hope alluded to earlier was partially fulfilled some time ago, when it was shown that the use of branch cut singularities could be a powerful tool in the calculations of scattering amplitudes. This program has given birth to a solid framework which nowadays is known as the unitarity based method~\cite{Bern:1994zx,Bern:1994cg,Bern:1995db,Bern:1996je,Bern:1996ja,Bern:1997sc,Bern:2004cz}. This technique has since streamlined the evaluation of many loop amplitudes; at one-loop, the problem of computing amplitudes boils down to computing tree amplitudes -- in many cases, we also need to calculate rational terms.

On the other hand, discontinuities in a given physical channel also possess by itself an elaborate analytic structure. Hence the cutting process can be repeated over and over again, which leads to what became known as generalized unitarity constraints~\cite{Britto:2004nc,Forde:2007mi,Kosower:12,Caron-Huot:2012awx,Johansson:2012zv,Johansson:2013sda,Abreu:2017ptx,Sogaard:2014jla,Larsen:2015ped,Ita:2015tya,Remiddi:2016gno,Primo:2016ebd,Frellesvig:2017aai,Zeng:2017ipr,Ellis:12,Frellesvig:Thesis,Brandhuber:05,Britto:2010xq,Carrasco:11,Abreu:2015zaa,Bern:2004ky,Britto:2008vq,Bern:11,Bern:2010qa,Drummond:2008bq,Engelund:2013fja,Elvang:2019twd,Bern:2021ppb,Bern:2020ikv,Primo:2016omk}~\footnote{For a much more extensive body of research on unitarity-based methods, please check references within~\cite{Bern:1994zx,Bern:1994cg,Bern:1995db,Bern:1996je,Bern:1996ja,Bern:1997sc,Bern:2004cz,Britto:2004nc,Forde:2007mi,Kosower:12,Caron-Huot:2012awx,Johansson:2012zv,Johansson:2013sda,Abreu:2017ptx,Sogaard:2014jla,Larsen:2015ped,Ita:2015tya,Remiddi:2016gno,Primo:2016ebd,Frellesvig:2017aai,Zeng:2017ipr,Ellis:12,Frellesvig:Thesis,Brandhuber:05,Britto:2010xq,Carrasco:11,Abreu:2015zaa,Bern:2004ky,Britto:2008vq,Bern:11,Bern:2010qa,Drummond:2008bq,Engelund:2013fja,Elvang:2019twd,Bern:2021ppb,Bern:2020ikv,Primo:2016omk}.}. It did not take long for many authors to realize that this could also be useful in the evaluation of loop amplitudes. In the method of generalized unitarity, the aim is to find an integrand that could reproduce all possible unitarity cuts. In the process one may require the integrand to reproduce each of the cut solutions independently. The resulting value for the cut calculated on each solution is called a leading singularity~\cite{Britto:2004nc,Arkani-Hamed:2008owk,Arkani-Hamed:2009kmp,Arkani-Hamed:2010zjl,book,Elvang:15,Buchbinder:2005wp,Cachazo:2008dx,Cachazo:2008vp}. This technique is reminiscent of the so-called maximal-cut method in the sense that it also requires the cutting of a maximal (or near-maximal) number of propagators. To put it simply, leading singularities are generalizations of standard unitarity cuts. The latter evaluates discontinuities across co-dimension one branch cuts; leading singularities are associated with singularities of the highest possible co-dimension and are evaluated as multidimensional residues, possessing support outside the physical region of integration. 
 
Leading singularities comprise a powerful technique to compute amplitudes. Whereas standard unitarity cuts might have divergences, leading singularities are calculated using compact contours and are hence finite. Moreover, leading singularities encompass physical states and are gauge invariant, a property also enjoyed by standard unitarity cuts. These features make them useful quantities to employ in the study of general theories. For instance, it was shown that it is also a valuable contrivance when addressing classical scattering and as a consequence it can be helpful in the calculation of many important classical observables, reducing many calculations which could be very complicated within the framework of general relativity~\cite{Cachazo:2017jef,Guevara:17,Guevara:19a,MS:22}. In this work we wish to show that they can also be useful in the study of analytic properties of scattering amplitudes in Lee-Wick-type theories. This kind of higher-derivative theory was shown to be unitary despite the presence of an unstable ghostlike resonance~\cite{DM:19}. We are particularly interested in a higher-derivative version of the Yang-Mills theory and also quadratic gravity. The latter still constitutes a potential UV completion for quantum gravity~\cite{Donoghue:2018izj,Donoghue:2021cza}. We will study one-loop scattering of gluons and gravitons as well as scattering of matter particles in such higher-derivative theories. We remark that leading singularities in in cubic theories of gravity were already considered in Refs.~\cite{Emond:2019crr,Burger:2019wkq}.

Even though unitarity in theories with unstable particles is only assured when unitarity cuts are not taken through unstable propagators~\cite{DM:19,Veltman:63,Veltman:74,Rodenburg,Lang,Denner:2014zga}, recently it was argued that the unitarity method still works for loop amplitudes containing resonances~\cite{Menezes:2021tsj}. This implies that the leading-singularity technique is also available to approach amplitudes in such theories. On the other hand, causality is violated on microscopic scales of order the width of the resonance~\cite{Lee:1969fy,Coleman:1969xz,Grinstein:08,Grinstein:2008bg,Donoghue:2019ecz,Donoghue:2020mdd,Donoghue:2021meq}. This is a consequence of the fact that the poles associated with such unusual resonances are on the physical sheet of the analytic continuation of the scattering amplitude, therefore the resummed propagators do not satisfy the usual analyticity properties. In the calculation of one-loop Feynman diagrams using standard methods (or when proving the cutting rules directly from Feynman diagrams), this demands the usage of a deformed contour, the Lee-Wick contour~\cite{Lee:1969fy,Grinstein:2008bg,DM:19}, perhaps together with some additional prescription, known in the literature as the Cutkosky-Landshoff-Olive-Polkinghorn (CLOP) prescription~\cite{Cutkosky:1969fq}. Even though important recent efforts were made towards a better understanding of analytic properties of amplitudes of higher-derivative theories~\cite{Aglietti:2016pwz,Anselmi:2017yux,Anselmi:2017lia,Anselmi:2018kgz,Anselmi:2018tmf,Anselmi:2021hab}, modern tools were put in an application only recently~\cite{Johansson:17,Johansson:18,Azevedo:2019zbn,Menezes:2021dyp}. Our investigation therefore furnishes a promising avenue for the comprehension of the analytic structure of loop amplitudes in higher-derivative Yang-Mills and quadratic gravity theories. Here we will use units such that $\hbar=c=1$. We take the Minkowski metric as $\eta_{\mu\nu} = \textrm{diag}(1,-1,-1,-1)$ and the Riemann curvature tensor given by $R^{\lambda}_{\ \mu\nu\kappa} = \partial_{\kappa}\Gamma^{\lambda}_{\mu\nu} + \Gamma^{\eta}_{\mu\nu}\Gamma^{\lambda}_{\kappa\eta} - (\nu \leftrightarrow \kappa)$.

\section{Brief review of color-kinematics duality and Yang-Mills amplitudes}

In this section we will give a brief review of an intriguing result involving Yang-Mills amplitudes which became known as the color-kinematics duality~\cite{CK,Bern:08,Johansson:15,Mastrolia:2015maa,Mafra:2011kj,Bjerrum-Bohr:2016axv,Du:2017kpo,Du:2012mt,Yuan:2012rg,Boels:2012ew,Boels:2013bi,Bjerrum-Bohr:2013iza,Bern:2013yya,Mogull:2015adi,He:2015wgf,Yang:2016ear,He:2017spx,Borsten:2021rmh,Chiodaroli:2013upa,Brandhuber:2021bsf,Chen:2013fya,Kampf:2021jvf,Du:2016tbc,Low:2019wuv}. We will also list all important gauge-theory amplitudes that will be important in the calculations that follow.

Tree-level amplitudes of Yang-Mills theory can be written as a sum over distinct trivalent diagrams~\cite{Elvang:15}
\begin{equation}
A_n = \sum_{k} \frac{c_k n_k}{s_k}
\end{equation}
where $s_k$ are inverse propagators (which could be massive), the $c_k$s are associated color factors and the numerators $n_k$ are in general polynomials of Lorentz-invariant contractions of polarization vectors and momenta. When the color factors satisfy Jacobi identities such as
\beq
c_i + c_j + c_k = 0
\eeq
one can easily prove that the amplitude is invariant under generalized gauge transformations:
\beq
n_i \to n_i + s_i \Delta,
\,\,\,
n_j \to n_j + s_j\Delta,
\,\,\,
n_k \to n_k + s_k \Delta,
\eeq
where $\Delta$ is an arbitrary function and $s_i, s_j, s_k$ are inverse propagators not shared among the different diagrams associated with each of the above color factors. On the other hand, this also implies that such numerators are not, in general, gauge invariant, despite the fact that the amplitude is. Moreover, the above color factors can also obey relations such as $c_i = - c_j$.

The so-called color-kinematics duality states that there exists at least one representation for gauge-theory amplitudes such that the numerators will satisfy identical algebraic properties as the corresponding color factors, that is
\beq
n_i + n_j + n_k = 0
\eeq
and also, for instance, $n_i = - n_j$. The fact that such gauge-dependent numerators obey these relations has one direct important consequence: One can derive gauge-invariant homogeneous relations between color-ordered partial amplitudes~\cite{Bern:08}. These are the so-called BCJ amplitude relations. 

A somewhat different perspective arises in gauge theories with higher-derivative operators in the Lagrangian. Here we will be particularly interested in the higher-derivative Yang-Mills with a Lagrangian density given by~\cite{Grinstein:08,Johansson:18,Menezes:2021dyp}
\begin{equation}
M^2 {\cal L} = - \frac{M^2}{4} F^{a}_{\mu\nu} F^{a \mu\nu} 
+ \frac{1}{2} D_{\mu} F^{a \mu\nu} D_{\lambda} F^{a \lambda}_{\ \ \nu}
\end{equation}
where covariant derivative in the adjoint representation reads
\begin{equation}
D^{\mu} F_{\alpha\beta}^a = \partial^{\mu} F_{\alpha\beta}^a + g  f^{abc} A^{b \mu} F_{\alpha\beta}^c .
\end{equation}
As well known, higher-derivative terms modifies the propagation of particles in such a way that the associated propagator develops a massive contribution with wrong-sign residue. We can interpret such a massive term as a new degree of freedom of the theory, a ghost ``particle" in itself, one carrying a different causal direction, as discussed in Refs.~\cite{Donoghue:2019ecz,Donoghue:2021meq}. Such ghost particles were dubbed {\it Merlin modes} elsewhere~\cite{Donoghue:2019ecz,Donoghue:2020mdd}. Even though Merlin modes could give rise to major problems, it was proved that the theory is unitary~\cite{DM:19} and possibly stable, at least in certain contexts~\cite{Donoghue:2021eto,Salvio:2019ewf}. These are consequences of the fact that the Merlin modes are unstable and hence they cannot appear in the asymptotic spectrum. The possible drawback is that such theories predict causality violation for microscopic time scales. However, as the theory possesses unitary time evolution, one cannot verify paradoxes in experiments that involve normal particles in the asymptotic states -- the theory may have unconventional acausal effects, but this feature does not seem to make it inconsistent.

We take the total gauge propagator in the form~\cite{Menezes:2021dyp}
\begin{equation}
D^{ab}_{\mu\nu}(p) = - \frac{\delta^{ab}}{p^2} \left( \eta_{\mu\nu} - (1-\xi) \frac{p_{\mu} p_{\nu}}{p^2} \right)
+ \frac{\delta^{ab}}{p^2 - M^2 - i M \Gamma} \left( \eta_{\mu\nu} - \frac{p_{\mu} p_{\nu}}{M^2} \right) .
\end{equation}
Henceforth we will consider the Feynman gauge, $\xi = 1$. The second contribution is the Merlin propagator in the narrow-width approximation, with $\Gamma \ll M$ being the decay width of the Merlin particle. Recall that, in general, one should consider a resummed form for unstable propagators as perturbation theory breaks down in the resonance region. Observe the two unusual minus signs in this equation; the change of the two signs together implies that the imaginary part of the Merlin propagator is the same as a normal resonance, a feature of utmost importance for unitarity to hold -- and, as a result, the generalized unitarity method still makes sense in the applications for higher-derivative theories. In what follows, unless otherwise stated, the contribution of the width $\Gamma$ will be left implicit in the Merlin propagators.

Concerning color-kinematics duality, it was shown in Ref.~\cite{Menezes:2021dyp} that $4$-point amplitudes in such a theory still obey the duality, but with an unusual feature -- the presence of quartic propagators renders the numerators in the amplitudes gauge invariant and as a consequence color-ordered amplitudes will not display conventional BCJ relations. In this case, the amplitude representation we have derived is necessarily unique and color-kinematics duality is trivially satisfied.

We will now list some of the tree-level amplitudes of higher-derivative Yang-Mills theories that will be used in the considerations that will follow. Let us begin with the $3$-particle amplitude. As demonstrated in Refs.~\cite{Johansson:18,Menezes:2021dyp}, $3$-particle amplitudes involving only physical gluons will not display contributions coming from higher-order derivative terms:
\begin{equation}
A^{(4)}_{3}[1^{h_1},2^{h_2},3^{h_3}] = M^2 A^{(2)}_{3}[1^{h_1},2^{h_2},3^{h_3}]
\label{g}
\end{equation}
where the $h_{i}$s are different helicities of the gluons and the superscripts represent the four- and two-derivative theories. Square brackets for color-ordered gauge-theory amplitudes are used in order to make an explicit distinction from non- color-ordered amplitudes. A coefficient $g /M^2$ was also factored out. In particular, such gluon amplitudes are completely fixed by little group scaling and locality. We find that~\cite{Elvang:15}
\bea
A^{(2)}_{3}[1^{-},2^{-},3^{+}] &=& \frac{\langle 12 \rangle^3}{\langle 13 \rangle \langle 32 \rangle}
\nn\\
A^{(2)}_{3}[1^{+},2^{+},3^{-}] &=& \frac{\bigl[ 12 \bigr]^3}{\bigl[ 13 \bigr] \bigl[ 32 \bigr]} .
\eea
By employing BCFW recursion relations~\cite{BCFW1,BCFW2}, one can show that the result~(\ref{g}) generalizes to an arbitrary number of gluons: 
\begin{equation}
A^{(4)}_{n}[1^{h_1},2^{h_2},3^{h_3},\ldots,n^{h_n}] = 
M^2 A^{(2)}_{n}[1^{h_1},2^{h_2},3^{h_3},\ldots,n^{h_n}]
\end{equation}
Furthermore, momentum conservation allows us to show that the $3$-particle on-shell amplitude involving a single Merlin particle vanishes~\cite{Johansson:18,Menezes:2021dyp}:
\begin{equation}
A^{(4)}_{3}(1^{h_1},2^{h_2}, 3^{IJ}) = 0,
\end{equation}
which also can be generalized to
\begin{equation}
A^{(4)}_{n+1}(1^{h_1},2^{h_2},\ldots,n^{h_n},k^{IJ}) = 0 .
\end{equation} 
Observe that massive gauge bosons carry explicit $SU(2)$ little-group indices. In what follows we will use instead a bold notation to suitably indicate symmetric combinations of $SU(2)$ little-group indices for massive spinors. For a pedagogical discussion of the formalism for massive spinors adopted in the present work, see Refs.~\cite{Aoude:19,Shadmi:19,Durieux:20,Arkani-Hamed:17,Chung:19}. 

Finally, the case with two Merlin particles is non-trivial; we find that~\cite{Menezes:2021dyp,Johansson:19}
\begin{eqnarray}
A_{3}[1^{+}, {\bf 2}, {\bf 3}] &=& \sqrt{2}  
\frac{\langle r| {\bf 3} | 1\bigr]}{\langle r1 \rangle} 
\langle {\bf 3} {\bf 2} \rangle^2
\nonumber\\
A_{3}[1^{-}, {\bf 2}, {\bf 3}] &=& \sqrt{2}  
\frac{\bigl[ r| {\bf 3} | 1\rangle}{\bigl[ 1r \bigr]} 
\bigl[ {\bf 3} {\bf 2} \bigr]^2 .
\end{eqnarray}
To complete our list of $3$-particle on-shell amplitudes, let us write down the results involving gluons or Merlins with two scalars. Actually the amplitudes have the same form -- the difference relies on the polarization vectors. For gluons, which are massless, the three-point amplitudes for one positive-helicity gluon and two massive (complex) scalars of mass $\mu^2$ reads~\cite{Henn:14}
\beq
A_{3}^{\textrm{tree}}(\ell_{i A}^{+},p^{a +},\ell_{j B}^{-}) = g T^{a}_{ij}
( \ell^{\mu}_{A} - \ell^{\mu}_{B} ) \epsilon_{\mu}(p)
= g T^{a}_{ij} \sqrt{2} \frac{\langle \xi | \ell_A | p \bigr]}{\langle \xi p \rangle}
\eeq
where $T^a$ are $SU(N)$ generators in the fundamental representation, which satisfy
\bea
[ T^a, T^b ] &=& i f^{abc} T^c
\nn\\
\textrm{Tr}[ T^a T^b ] &=& \frac{1}{2} \delta^{ab}
\eea
($f^{abc}$ are the structure constants) and $| \xi \rangle$ is an arbitrary reference spinor not proportional to $| p \rangle$. Now, for Merlins, which are massive, we obtain that~\cite{Menezes:2021dyp}
\begin{equation}
A_{3}^{\textrm{tree}}(\ell_{i A}^{+},{\bf p}^{a},\ell_{j B}^{-}) 
= g T^{a}_{ij}  ( \ell^{\mu}_{A} - \ell^{\mu}_{B} ) 
\epsilon_{\mu}({\bf p})
= g T^{a}_{ij}  \sqrt{2} \frac{ \langle {\bf p} | \ell_A | {\bf p} \bigr]  }{M} .
\end{equation}
Let us now move on to $4$-point amplitudes. In the situation involving only gluons, we have already noted that the amplitudes are essentially the same as in the pure Yang-Mills theory. In this case, nonvanishing color-ordered amplitudes are the so-called $N^{K}$MHV amplitudes which contain $K+2$ negative-helicity gluons and $n-K-2$ positive helicity gluons. The MHV and the antiMHV amplitudes are given by the Parke-Taylor formula~\cite{PT:86}
\bea
A^{\textrm{MHV,tree}}_{n}[1^{+}, \cdots, i^{-}, \cdots, j^{-}, \cdots, n^{+}] &=& \frac{\langle ij \rangle^4}
{\prod_{i=1}^n \langle i, i+1 \rangle}
\nn\\
A^{\textrm{antiMHV,tree}}_{n}[1^{-}, \cdots, i^{+}, \cdots, j^{+}, \cdots, n^{-}] &=& \frac{\bigl[ ij \bigr]^4}
{\prod_{i=1}^n \bigl[ i, i+1 \bigr]}
\eea
respectively, where $n+1 \to 1$. The Parke-Taylor formula can be proved using the BCFW recursion relations. 

Compton amplitudes involving Merlin particles were considered in detail in Ref.~\cite{Menezes:2021dyp}. In particular, we are interested in the color-ordered amplitude $A_{4}[{\bf 2}, 1^{+},4^{+}, {\bf 3}]$ with Merlins and gluons. Using BCFW recursion relations, one finds that~\cite{Menezes:2021dyp,Johansson:19}
\begin{eqnarray}
A_{4}[{\bf 2},1^{+},4^{+},  {\bf 3}] &=& 2 M^4 \frac{\bigl[ 14 \bigr]}{\langle 14 \rangle}
\frac{\langle {\bf 3} {\bf 2} \rangle^2}{(p_1 + p_2)^2 - M^2} .
\end{eqnarray}
The other Compton scattering amplitudes that will be used in the calculation of the associated leading singularities are the ones with two gauge particles and two massive fundamental scalars with mass $m$. The ones with gluons read~\cite{Arkani-Hamed:17,Johansson:19,Plefka:2019wyg}
\bea
A_{4}(\ell_{i A},1^{a +}, 2^{b -}, \ell_{j B}) &=& 
- \frac{2 g^2}{t}  \left( \frac{T^{a}_{ik} T^{b}_{kj}}{2 \ell_A \cdot p_1} 
+ \frac{T^{b}_{ik} T^{a}_{kj}}{2 \ell_A \cdot p_2} \right)
\langle 2 | \ell_A | 1 \bigl]^{2}
\nn\\
A_{4}(\ell_{i A},1^{a -}, 2^{b -}, \ell_{j B}) &=& 2 m^2 g^2 
\frac{\langle 2 1 \rangle^2}{(p_1 + p_2)^2} 
\left[\frac{T^{a}_{ik} T^{b}_{lk}}{(\ell_A + p_1)^2 - m^2} 
+\frac{T^{b}_{ik} T^{a}_{kj}}{(\ell_A + p_2)^2 - m^2} \right] .
\eea
The associated reverse-helicity amplitudes can be obtained from those above by swapping angle and square brackets. In turn, the amplitude with external Merlins is given by~\cite{Menezes:2021dyp}
\bea
A_{4}(\ell_{i A}, {\bf 1}^a, {\bf 2}^b, \ell_{j B}) &=& 2 g^2 T^{a}_{ik} T^{b}_{kj}
\frac{ \langle {\bf 1} | \ell_A | {\bf 1} \bigr]  \langle {\bf 2} | \ell_B | {\bf 2} \bigr] }{M^2} 
\frac{1}{(\ell_A + p_1)^2 - m^2}
+ 2 g^2 T^{b}_{ik} T^{a}_{kj}
\frac{ \langle {\bf 1} | \ell_B | {\bf 1} \bigr] \langle {\bf 2} | \ell_A | {\bf 2} \bigr] }{M^2} 
\frac{1}{(\ell_A + p_2)^2 - m^2}
\nn\\
&-& i g^2 f^{bac} T^{c}_{ij} \frac{2}{M^2} \left[ \langle {\bf 1} {\bf 2} \rangle \bigl[ {\bf 2} {\bf 1} \bigr] ( {\bf 1} - {\bf 2} ) \cdot \ell_A 
-  \Bigl( \langle {\bf 1} | \ell_{A} | {\bf 1} \bigr] \langle {\bf 2} | {\bf 1} | {\bf 2} \bigr] 
- \langle {\bf 2} | \ell_{A} | {\bf 2} \bigr] \langle {\bf 1} | {\bf 2} | {\bf 1} \bigr] \Bigr)
\right]
\nn\\
&\times& \left( \frac{ 1 }{(p_1 + p_2)^2} - \frac{ 2 }{(p_1 + p_2)^2 - M^2} \right)
\nn\\
&+& g^2  \{ T^{a}, T^{b} \}_{ij} \frac{ \langle {\bf 1} {\bf 2} \rangle \bigl[ {\bf 2} {\bf 1} \bigr]  }{M^2} .
\eea
The overall minus sign in front of the last term in the $t$-chanell is a consequence of the ghost feature of the Merlin mode.

There is still one amplitude we need that was not calculated previously -- this is the $4$-point amplitude with scalars, a Merlin and a gluon as external states. In order to build up such a scattering amplitude, we will consider the contributions due to consistent factorization in all possible channels given the three-particle amplitudes. The residue in the $s$-channel is given by
\beq
\textrm{Res}_{s} = 2 \sqrt{2} \frac{ \langle {\bf 1} | \ell_A | {\bf 1} \bigr]  }{M}
\ell^{\mu}_{B} \epsilon_{\mu}(2) .
\eeq
The residue in the $u$-channel is obtained by swapping the Merlin with the gluon. As amplitudes involving a single Merlin particle vanish, the residue in the $t$-channel presents us with only one possibility -- the pole must come from the massive Merlin particle. So the residue is given by the product of a $3$-point amplitude involving two scalars and a Merlin and with another $3$-point amplitude with two Merlins and a gluon~\footnote{For this calculation one can make use of the triple-gauge vertex calculated in detail in  Ref.~\cite{Menezes:2021dyp}}:
\bea
\textrm{Res}_{t} &=&
- M^2
\Bigl[ \eta_{\mu\nu} ( p_{1}  - p_2 )_{\rho} + \eta_{\nu\rho} ( 2 p_{2} + p_{1} )_{\mu}
- \eta_{\rho\mu} ( 2 p_{1} + p_2 )_{\nu} \Bigr]
\epsilon^{\mu IJ}({\bf 1}) \epsilon^{\nu}(2) \epsilon^{\rho MN}({\bf p})
2 \ell_{A {\alpha}}  [\epsilon^{ \alpha}_{MN}({\bf p})]^{*}
\nn\\
&=&
2 M^2
\Bigl[ \eta_{\mu\nu} ( p_{1}  - p_2 )_{\rho} + \eta_{\nu\rho} ( 2 p_{2} + p_{1} )_{\mu}
- \eta_{\rho\mu} ( 2 p_{1} + p_2 )_{\nu} \Bigr]
\epsilon^{\mu IJ}({\bf 1}) \epsilon^{\nu}(2)  \ell_{A {\alpha}}
\left( \eta^{\rho\alpha} - \frac{p^{\rho} p^{\alpha}}{M^2} \right)
\nn\\
&=& 2 M^2
\Bigl[  ( \epsilon({\bf 1}) \cdot \epsilon(2) )  ( p_1  - p_2 ) \cdot \ell_{A}
- 2 \Bigl( \ell_{A} \cdot \epsilon({\bf 1}) p_1 \cdot \epsilon(2)
- \ell_{A} \cdot \epsilon(2) p_2 \cdot \epsilon({\bf 1})  \Bigr) \Bigr]
+ M^4 \epsilon({\bf 1}) \cdot \epsilon(2)
\eea
where we used that $2 \ell_A \cdot {\bf p} = M^2$. We also have a contact term:
\beq
\textrm{Contact-Term} \, [{\bf 1}, 2] = \epsilon({\bf 1}) \cdot \epsilon(2)  .
\eeq
Collecting our results, restoring color factors and the factor $g/M^2$ in the $t$-channel: 
\bea
A_{4}(\ell_{i A}, {\bf 1}^a, 2^b, \ell_{j B}) &=&
2 \sqrt{2} g^2 T^{a}_{ik} T^{b}_{kj} \frac{ \langle {\bf 1} | \ell_A | {\bf 1} \bigr]  }{M}
 \frac{\ell^{\mu}_{B} \epsilon_{\mu}(2)}{(\ell_A + {\bf 1})^2 - m^2}
+ 2 \sqrt{2} g^2 T^{b}_{ik} T^{a}_{kj} \frac{ \langle {\bf 1} | \ell_B | {\bf 1} \bigr]  }{M}
 \frac{\ell^{\mu}_{A} \epsilon_{\mu}(2)}{(\ell_B + {\bf 1})^2 - m^2}
\nn\\
&-& i f^{abc} T^c_{ij} g^2 \Bigl\{ 2 
\Bigl[  ( \epsilon({\bf 1}) \cdot \epsilon(2) )  ( p_1  - p_2 ) \cdot \ell_{A}
- 2 \Bigl( \ell_{A} \cdot \epsilon({\bf 1}) p_1 \cdot \epsilon(2)
- \ell_{A} \cdot \epsilon(2) p_2 \cdot \epsilon({\bf 1})  \Bigr) \Bigr]
+ M^2 \epsilon({\bf 1}) \cdot \epsilon(2) \Bigr\}
\nn\\
&\times& \frac{1}{(p_1 + p_2)^2 - M^2}
\nn\\
&+& g^2 \{ T^a,T^b \}_{ij} \epsilon({\bf 1}) \cdot \epsilon(2) .
\eea
Observe that, swapping the labels ``gluon" and ``Merlin" results in a change of sign of the last term in the $t$-channel:
\bea
A_{4}(\ell_{i A}, 1^a, {\bf 2}^b, \ell_{j B}) &=&
2 \sqrt{2} g^2 T^{a}_{ik} T^{b}_{kj} \frac{ \langle {\bf 2} | \ell_B | {\bf 2} \bigr]  }{M}
\ell^{\mu}_{A} \epsilon_{\mu}(1) \frac{1}{(\ell_B + {\bf 2})^2 - m^2}
\nn\\
&+& 
2 \sqrt{2} g^2 T^{b}_{ik} T^{a}_{kj} \frac{ \langle {\bf 2} | \ell_A | {\bf 2} \bigr]  }{M}
\ell^{\mu}_{B} \epsilon_{\mu}(1) \frac{1}{(\ell_A + {\bf 2})^2 - m^2}
\nn\\
&-& i f^{abc} T^c_{ij} g^2 \Bigl\{ 2 
\Bigl[  ( \epsilon(1) \cdot \epsilon({\bf 2}) )  ( p_1  - p_2 ) \cdot \ell_{A}
- 2 \Bigl( \ell_{A} \cdot \epsilon(1) p_1 \cdot \epsilon({\bf 2})
- \ell_{A} \cdot \epsilon({\bf 2}) p_2 \cdot \epsilon(1)  \Bigr) \Bigr]
- M^2 \epsilon(1) \cdot \epsilon({\bf 2}) \Bigr\}
\nn\\
&\times& \frac{1}{(p_1 + p_2)^2 - M^2}
\nn\\
&+& g^2 \{ T^b,T^a \}_{ij} \epsilon(1) \cdot \epsilon({\bf 2}) .
\eea
Let us verify whether the above amplitude satisfies color-kinematics duality. We write:
\beq
A_{4}(\ell_{i A}, {\bf 1}^a, 2^b, \ell_{j B}) = \frac{c_s n_s}{(\ell_A + p_1)^2 - m^2}
+  \frac{c_u n_u}{(\ell_A + p_2)^2 - m^2} - \frac{c_t n_t}{(p_1+p_2)^2-M^2}
\eeq
where we have defined the following numerators
\bea
n_s &=& 2 \sqrt{2} g^2 \frac{ \langle {\bf 1} | \ell_A | {\bf 1} \bigr]  }{M} \ell^{\mu}_{B} \epsilon_{\mu}(2)
+ g^2 \epsilon({\bf 1}) \cdot \epsilon(2) \Bigl( 2 \ell_A \cdot p_1 + M^2 \Bigr)
\nn\\
n_u &=& 2 \sqrt{2} g^2 \frac{ \langle {\bf 1} | \ell_B | {\bf 1} \bigr]  }{M}\ell^{\mu}_{A} \epsilon_{\mu}(2)
+ g^2 \epsilon({\bf 1}) \cdot \epsilon(2) \Bigl( 2 \ell_A \cdot p_2 \Bigr)
\nn\\
n_t &=& g^2 \Bigl\{ 2 \Bigl[  ( \epsilon({\bf 1}) \cdot \epsilon(2) )  ( p_1  - p_2 ) \cdot \ell_{A}
- 2 \Bigl( \ell_{A} \cdot \epsilon({\bf 1}) p_1 \cdot \epsilon(2)
- \ell_{A} \cdot \epsilon(2) p_2 \cdot \epsilon({\bf 1})  \Bigr) \Bigr]
+ M^2 \epsilon({\bf 1}) \cdot \epsilon(2) \Bigr\}
\nn\\
\eea
and color factors
\bea
c_s &=& T^{a}_{ik} T^{b}_{kj}
\nn\\
c_u &=& T^{b}_{ik} T^{a}_{kj}
\nn\\
c_t &=& i f^{abc} T^c_{ij} = [T^a,T^b]_{ij} .
\eea
Observe that $c_t$ is antisymmetric under an interchange of two legs, and the corresponding numerator $n_t$ also obeys this property (to see this for the last term, one must check the construction of the amplitude given above). Moreover, $c_s - c_u = c_t$ implies that $n_s - n_u = n_t$. So we have obtained a representation for the amplitude that satisfies color-kinematics duality. A similar result is obtained for the other amplitude, that is:
\beq
A_{4}(\ell_{i A}, 1^a, {\bf 2}^b, \ell_{j B}) = \frac{c_s n_s}{(\ell_A + p_2)^2 - m^2}
+  \frac{c_u n_u}{(\ell_A + p_1)^2 - m^2} - \frac{c_t n_t}{(p_1+p_2)^2-M^2}
\eeq
with
\bea
n_s &=& 2 \sqrt{2} g^2  \frac{ \langle {\bf 2} | \ell_B | {\bf 2} \bigr]  }{M} \ell^{\mu}_{A} \epsilon_{\mu}(1)
+ g^2 \epsilon(1) \cdot \epsilon({\bf 2})  \Bigl( 2 \ell_A \cdot p_1 \Bigr)
\nn\\
n_u &=& 2 \sqrt{2} g^2 \frac{ \langle {\bf 2} | \ell_A | {\bf 2} \bigr]  }{M} \ell^{\mu}_{B} \epsilon_{\mu}(1) 
+ g^2 \epsilon(1) \cdot \epsilon({\bf 2})\Bigl( 2 \ell_A \cdot p_2 + M^2 \Bigr)
\nn\\
n_t &=& g^2 \Bigl\{ 2 
\Bigl[  ( \epsilon(1) \cdot \epsilon({\bf 2}) )  ( p_1  - p_2 ) \cdot \ell_{A}
- 2 \Bigl( \ell_{A} \cdot \epsilon(1) p_1 \cdot \epsilon({\bf 2})
- \ell_{A} \cdot \epsilon({\bf 2}) p_2 \cdot \epsilon(1)  \Bigr) \Bigr]
- M^2 \epsilon(1) \cdot \epsilon({\bf 2}) \Bigr\}
\nn\\
\eea
and
\bea
c_s &=& T^{a}_{ik} T^{b}_{kj}
\nn\\
c_u &=& T^{b}_{ik} T^{a}_{kj}
\nn\\
c_t &=& i f^{abc} T^c_{ij} = [T^a,T^b]_{ij} .
\eea

\section{Brief review of the double-copy method and gravity amplitudes}

As a consequence of the color-kinematics duality, one can derive amplitudes for a gravity theory with a spectrum which is the square of the Yang-Mills spectrum. One just need to replace color factors for kinematic factors in the corresponding gauge-theory amplitude~\cite{Plefka:2019wyg,Bern:100,Bern:10,Bern:2015ooa,Bern:2017yxu,Carrasco:2020ywq,Oxburgh:2012zr,Ochirov:2013xba,Borsten:2020zgj,Low:2020ubn,Brandhuber:21,Johnson:20,Chiodaroli:2017ngp}:
\begin{equation}
M_n = \sum_{k} \frac{n_k n_k}{s_k}.
\end{equation}
We have suppressed a factor of the coupling $(\kappa/2)^{n-2}$ at $n$ points, where $\kappa^2 = 32 \pi G$. This is the BCJ double-copy relation. The squaring relation can be generalized to
\begin{equation}
M_n = \sum_{k} \frac{\widetilde{n}_k n_k}{s_k}
\end{equation}
where $\widetilde{n}_k$ and $n_k$ belong to two distinct Yang-Mills numerators, with only one of them constrained to satisfy color-kinematics duality.

One can also use the color-kinematics relation together with the BCJ relations to rewrite gravity amplitudes as products of partial Yang-Mills amplitudes. The set of relations obtained in this way are called
Kawai-Lewellen-Tye (KLT) relations~\cite{KLT}. The KLT relations streamlines the usual cumbersome calculations found when dealing with gravitational interactions: Gravity tree amplitudes can be written in terms of gauge-theory partial amplitudes. Through four points these relations are given by
\begin{eqnarray}
M^{\textrm{tree}}_{3}(1,2,3) &=& i A^{\textrm{tree}}_{3}[1,2,3] A^{\textrm{tree}}_{3}[1,2,3]
\nonumber\\
M^{\textrm{tree}}_{4}(1,2,3,4) &=& - i s_{12} A^{\textrm{tree}}_{4}[1,2,3,4] A^{\textrm{tree}}_{4}[1,2,4,3] 
\end{eqnarray}
where $M^{\textrm{tree}}_n$ are tree level gravity amplitudes. The basic building block is given by gauge-theory tree-level scattering amplitudes. 

The KLT relations are usually constructed for massless amplitudes, but a double-copy prescription for Weyl gravity was put forward in Refs.~\cite{Johansson:17,Johansson:18,Menezes:2021dyp}. Similar to the gauge-theory case, we also verify an unusual feature: Since standard BCJ relations are not obeyed by $4$-point gauge amplitudes, the corresponding KLT relations are not satisfied by $4$-point quadratic-gravity amplitudes. This feature was duly explored in Ref.~\cite{Menezes:2021dyp} in order to build up a consistent map to relate gauge amplitudes with gravity amplitude through a double-copy prescription This map is given by
\begin{equation}
(\textrm{Higher-derivative YM}) \otimes \textrm{YM}
= \textrm{Weyl-Einstein}
\label{doublecopy}
\end{equation}
which was motivated by the considerations studied in Refs.~\cite{Johansson:17,Johansson:18}. The action that describes the Weyl-Einstein theory reads
\begin{equation}
S = \int d^4x \sqrt{-g}
\left[\frac{2}{\kappa^2} R 
- \frac{1}{2\xi^2} C_{\mu\nu\alpha\beta}C^{\mu\nu\alpha\beta} \right]
\label{einsteinf}
\end{equation}
where $C_{\mu\nu\alpha\beta}$ is the Weyl tensor. This action also describes the more general quadratic-gravity theory in the Einstein frame. This can be obtained through a simple field redefinition from the Jordan-frame action~\cite{Salvio:18}
\begin{equation}
S = \int d^4x \sqrt{-g}
\left[\frac{2}{\kappa^2} R + \frac{1}{6 f_0^2} R^2 - \frac{1}{2\xi^2} C_{\mu\nu\alpha\beta}C^{\mu\nu\alpha\beta} \right].
\end{equation}
Essentially the difference between the two frames is that in the former we verify the presence of an extra scalar field with a particular interaction potential whose contribution is usually added to the matter sector. 

As extensively discussed in Ref.~\cite{Menezes:2021dyp}, the map given by Eq.~(\ref{doublecopy}) does not yield a pure Weyl-Einstein gravity. Typically the double copy of two vector fields produces gravitons, a dilaton and a axion. For quadratic gravity, the map Eq.~(\ref{doublecopy}) produces, besides such degrees of freedom, the associated Merlins with each of those particles. The projection to pure gravity can be obtained by correlating the helicities in the two gauge-theory copies (for the graviton) and by taking the symmetric tensor product of gauge-theory Merlins (for the gravitational Merlin). On the other hand, the map~(\ref{doublecopy}) requires, for consistency, the consideration of spontaneously broken gauge theories. In other words, the pure YM part of the double copy must comprise massive gauge fields. For a thorough discussion, we refer the reader the Ref.~\cite{Menezes:2021dyp}. For an interesting account of the double-copy construction for spontaneously broken gauge theories, see Refs.~\cite{Chiodaroli:17,Chiodaroli:18,Chiodaroli:19}.

The polarization tensors of massive Merlin particles can be constructed as tensor products of the spin-$1$ polarization vectors~\cite{Guevara:19a}:
\begin{equation}
\bigl[ \epsilon^{\alpha \dot{\alpha}} \bigr]^{(S)} = \bigl[ \epsilon^{\alpha \dot{\alpha}} \bigr]^{\otimes S}
= \left( \frac{ \sqrt{2} }{M} \right)^{S} \Bigl( | {\bf p} \rangle \bigl[ {\bf p} | \Bigr)^{S} 
\end{equation}
where symmetrization in the spinor products is implicitly understood.

Let us now list the on-shell amplitudes that will enter in the forthcoming calculations. As above, we begin with $3$-point amplitudes. $3$-particle amplitudes involving only physical gravitons do not display contributions coming from higher-order derivative terms~\cite{Johansson:18,Dona:2015tra,Menezes:2021dyp}, 
\begin{equation}
M^{(4)}_{3}[1^{h_1},2^{h_2},3^{h_3}] = M^2 M^{(2)}_{3}[1^{h_1},2^{h_2},3^{h_3}]
\end{equation}
($h_i$ represents a given graviton helicity) where
\bea
M^{(2)}_{3}(1^{-},2^{-},3^{+}) &=& \Bigl( A^{(2)}_{3}[1^{-},2^{-},3^{+}] \Bigr)^2 
= \frac{\langle 12 \rangle^6}{\langle 13 \rangle^2 \langle 32 \rangle^2}
\nn\\
M^{(2)}_{3}[1^{+},2^{+},3^{-}]  &=& \Bigl( A^{(2)}_{3}[1^{+},2^{+},3^{-}] \Bigr) 
= \frac{\bigl[ 12 \bigr]^6}{\bigl[ 13 \bigr]^2 \bigl[ 32 \bigr]^2} .
\eea
This result generalizes to an arbitrary number of gravitons by employing BCFW recursion relations. In turn, amplitudes with a single gravitational Merlin particle vanishes~\cite{Johansson:18,Menezes:2021dyp}:  
\begin{equation}
M^{(4)}_{n+1}(1^{h_1},2^{h_2},\ldots,n^{h_n},{\bf k}) = 0
\end{equation} 
The $3$-particle amplitude involving two gravitational Merlin particles and one graviton may be obtained from the YM case by using the KLT relations discussed in Ref.~\cite{Johansson:19}:
\begin{eqnarray}
M_{3}(1^{++},{\bf 2},{\bf 3}) &=& i A^{\textrm{tree, HD}}_{3}[1^{+},{\bf 2},{\bf 3}] 
A^{\textrm{tree,YM}}_{3}[1^{+},{\bf 2},{\bf 3}] = - 2i
\frac{\langle r| {\bf 3} | 1\bigr]^2}{M^4 \langle r1 \rangle^2} 
\langle {\bf 3} {\bf 2} \rangle^4
\nonumber\\
M_{3}(1^{--},{\bf 2},{\bf 3}) &=& i A^{\textrm{tree, HD}}_{3}[1^{-},{\bf 2},{\bf 3}] 
A^{\textrm{tree,YM}}_{3}[1^{-},{\bf 2},{\bf 3}] = - 2i \frac{\bigl[ r| {\bf 3} | 1\rangle^2}{M^4 \bigl[ 1r \bigr]^2} 
\bigl[ {\bf 3} {\bf 2} \bigr]^4 .
\end{eqnarray}
To complete our list of required $3$-particle amplitudes, we need the ones involving one gravitational particle (graviton or Merlin) and two massive scalars. The double-copy prescription was given in Ref.~\cite{Johansson:19}. The graviton case reads
\begin{eqnarray}
M_{3}^{\textrm{tree}}(\ell_A^{+},p^{++},\ell_B^{-})  
&=& i \Bigl( A^{\textrm{tree, YM}}_{3}[\ell_A^{+},p^{+},\ell_B^{-}] \Bigr)^2
\nonumber\\
&=& 2 i \frac{\langle \xi | \ell_A | p \bigr]^2}{\langle \xi p \rangle^2}
\end{eqnarray}
whereas in the Merlin case the amplitude is given by~\cite{Menezes:2021dyp}
\begin{eqnarray}
M_{3}^{\textrm{tree}}(\ell_A^{+},{\bf p},\ell_B^{-})  
&=&  i A^{\textrm{tree, HD}}_{3}[\ell_A^{+},{\bf p},\ell_B^{-}] 
A^{\textrm{tree,YM}}_{3}[\ell_A^{+},{\bf p},\ell_B^{-}]
\nonumber\\
&=& 2 i \frac{ \langle {\bf p} | \ell_A | {\bf p} \bigr]^2  }{M^2} .
\end{eqnarray}
Let us now display the $4$-point amplitudes that to go into the formulas of the leading singularities below. Amplitudes involving only gravitons can be obtained from the usual KLT relations. Likewise, amplitudes with gravitational Merlin particles and gravitons can also be derived with the same procedure~\cite{Menezes:2021dyp}. For instance, the amplitude $M_{4}[1^{++}, {\bf 2}, {\bf 3},4^{++}]$ which involves two gravitational Merlin particles reads~\cite{Johansson:19,Menezes:2021dyp}
\begin{equation}
M^{\textrm{tree}}_{4}({\bf 2},1^{++},4^{++},{\bf 3}) = - i s_{23} 
A^{\textrm{tree, HD}}_{4}[{\bf 2},1^{+},4^{+},{\bf 3}] A^{\textrm{tree, YM}}_{4}[{\bf 2},4^{+},1^{+},{\bf 3}] 
= 4i   \frac{\bigl[ 14 \bigr]^4}{s_{23}}
\frac{\langle {\bf 3} {\bf 2} \rangle^4}{(s_{12} - M^2)(s_{13} - M^2)} .
\end{equation}
where $s_{ij} = (p_i + p_j)^2$. As for the amplitudes involving scalar matter particles, we find that~\cite{Arkani-Hamed:17,Johansson:19}
\bea
M_{4}({\bf 1},2^{++}, 3^{--}, {\bf 4}) &=& -i  
 \frac{\langle 3| {\bf 1} |2 \bigl]^{4}}{[(p_1 + p_2)^2 - m^2] [(p_1 + p_3)^2 - m^2] (p_2 + p_3)^2}
\nn\\
M_{4}({\bf 1},2^{--}, 3^{--}, {\bf 4}) &=& - i m^{4}  
\frac{\langle 23 \rangle^4 }{(p_2 + p_3)^2}
\frac{1}{[(p_1 + p_2)^2 - m^2] [(p_1 + p_3)^2 - m^2]} 
\eea
for gravitons (reverse helicity amplitudes can be obtained from the above formulas by swapping angle and square brackets) and
\begin{eqnarray}
\hspace{-5mm}
M_{4}(\ell_A, {\bf 1}, {\bf 2}, \ell_B) &=& 
\left[ \frac{ 2 }{M^2} \langle {\bf 1} | \ell_A | {\bf 1} \bigr]  
\langle {\bf 2} | \ell_B | {\bf 2} \bigr]
+ \frac{\langle {\bf 1} {\bf 2} \rangle \bigl[ {\bf 2} {\bf 1} \bigr] }{M^2} (2 ({\bf 1} \cdot \ell_A) + M^2) \right]^2
 \frac{i}{(\ell_A + p_1)^2 - m^2}
\nonumber\\
&+&
\left[ \frac{2}{M^2}  \langle {\bf 1} | \ell_B | {\bf 1} \bigr] 
\langle {\bf 2} | \ell_A | {\bf 2} \bigr]
+ \frac{\langle {\bf 1} {\bf 2} \rangle \bigl[ {\bf 2} {\bf 1} \bigr] }{M^2} (2 ({\bf 2} \cdot \ell_A) + M^2) \right]^2
\frac{i}{(\ell_A + p_2)^2 - m^2}
\nonumber\\
&-& \frac{4 i}{M^4} 
\Bigl( \langle {\bf 1} {\bf 2} \rangle \bigl[ {\bf 2} {\bf 1} \bigr] ( {\bf 1} - {\bf 2} ) \cdot \ell_A 
+ \langle {\bf 1} | \ell_{A} | {\bf 1} \bigr] \langle {\bf 2} | \ell_B | {\bf 2} \bigr] 
- \langle {\bf 2} | \ell_{A} | {\bf 2} \bigr] \langle {\bf 1} | \ell_B | {\bf 1} \bigr] 
\Bigr)^2
\left( \frac{ 1 }{(p_1 + p_2)^2} - \frac{ 2 }{(p_1 + p_2)^2 - M^2} \right) .
\nn\\
\label{Mampscalar}
\end{eqnarray}
for Merlins~\cite{Menezes:2021dyp}. The former can be derived from the double-copy formula
\beq
M^{\textrm{tree}}(1_{s},2,3,4_{s}) = i \left( \frac{\kappa}{2} \right)^2 s_{23} 
(-1)^{ \left \lfloor{s}\right \rfloor - \left \lfloor{s_1}\right \rfloor - \left \lfloor{s_2}\right \rfloor + 1}
A^{\textrm{tree}}[1_{s_1},2,3,4_{s_1}] 
A^{\textrm{tree}}[1_{s_2},3,2,4_{s_2}] , \,\,\,
s = s_1 + s_2 
\label{DC}
\eeq
as proposed in Ref.~\cite{Johansson:19}. The amplitude with Merlins can be derived from the relation~\cite{Menezes:2021dyp}
\begin{equation}
M(1_s,2,3,4_s) = i \sum_{k} \frac{n^{(s_1)}_k(1_{s_1},2,3,4_{s_1}) \tilde{n}^{(s_2)}_k(1_{s_2},2,3,4_{s_2}) }{s_k}
\label{DCC}
\end{equation}
where $s=s_1+s_2$, $s_1, s_2$ are the associated spins of the matter particles, $2,3$ are graviton or Merlin particles, $\tilde{n}_k$ are numerators from the spontaneously broken gauge theory of the double copy and $s_k$ are inverse propagators. The double copy relation~(\ref{DCC}) implies that Eq.~(\ref{Mampscalar}) necessarily contains all the degrees of freedom produced by the double copy in the $t$-channel. In order to allow for only gravitons and gravitational Merlins to flow through the cuts of the $t$-channel, one may employ four-dimensional physical state projectors, defined as
\begin{eqnarray}
\sum_{\lambda = \pm 2} \epsilon^{\mu\nu}_{\lambda}(p;r) \epsilon^{\rho\sigma *}_{\lambda}(p;r)&=&
\frac{1}{2} \Bigl( \pi_{\mu\rho} \pi_{\nu\sigma} + \pi_{\mu\sigma} \pi_{\nu\rho} 
- \pi_{\mu\nu} \pi_{\rho\sigma} \Bigr) \,\,\,\ (\textrm{Graviton})
\nn\\
\pi_{\mu\nu} &=& \eta_{\mu\nu} - \frac{p_{\mu} \bar{p}_{\nu} + p_{\nu} \bar{p}_{\mu} }{p \cdot \bar{p}}
\nonumber\\
\sum_{\lambda = -2}^{2} \epsilon^{\mu\nu}_{\lambda} \epsilon^{\rho\sigma *}_{\lambda} &=&
\frac{1}{2} \Bigl( \widetilde{\pi}_{\mu\rho} \widetilde{\pi}_{\nu\sigma} 
+ \widetilde{\pi}_{\mu\sigma} \widetilde{\pi}_{\nu\rho} 
- \frac{2}{3} \widetilde{\pi}_{\mu\nu} \widetilde{\pi}_{\rho\sigma} \Bigr) \,\,\,\ (\textrm{Merlin})
\nn\\
\widetilde{\pi}_{\mu\nu} &=& \eta_{\mu\nu} - \frac{p_{\mu} p_{\nu} }{M^2} .
\label{projectors}
\end{eqnarray}
where $\bar{p}^{\mu} = (p^{0},-{\bf p})$. 

As in the previous case, our computations will require a $4$-point amplitude with a Merlin and a graviton as external states. Since the associated gauge-theory amplitude obeys color-kinematics duality, the evaluation of the corresponding gravity amplitude goes through the use of the double-copy prescription as given by Eq.~(\ref{DCC}). We obtain
\bea
&& M_{4}(\ell_{A}, {\bf 1}, 2, \ell_{B}) 
\nn\\
&& = 
\left[ 2 \sqrt{2} \frac{ \langle {\bf 1} | \ell_A | {\bf 1} \bigr]  }{M} \ell^{\mu}_{B} \epsilon_{\mu}(2)
+ \epsilon({\bf 1}) \cdot \epsilon(2) \Bigl( 2 \ell_A \cdot p_1 + M^2 \Bigr) \right]^2
\frac{i}{(\ell_A + p_1)^2 - m^2}
\nn\\
&& + 
\left[ 2 \sqrt{2} \frac{ \langle {\bf 1} | \ell_B | {\bf 1} \bigr]  }{M}\ell^{\mu}_{A} \epsilon_{\mu}(2)
+ g^2 \epsilon({\bf 1}) \cdot \epsilon(2) \Bigl( 2 \ell_A \cdot p_2 \Bigr) \right]^2
\frac{i}{(\ell_A + p_2)^2 - m^2} 
\nn\\
&& + 
\Bigl\{ 2 \Bigl[  ( \epsilon({\bf 1}) \cdot \epsilon(2) )  ( p_1  - p_2 ) \cdot \ell_{A}
- 2 \Bigl( \ell_{A} \cdot \epsilon({\bf 1}) p_1 \cdot \epsilon(2)
- \ell_{A} \cdot \epsilon(2) p_2 \cdot \epsilon({\bf 1})  \Bigr) \Bigr]
+ M^2 \epsilon({\bf 1}) \cdot \epsilon(2) \Bigr\}^2
\frac{i}{(p_1+p_2)^2-M^2}
\nn\\
\eea
and also
\bea
&& M_{4}(\ell_{A}, 1, {\bf 2}, \ell_{B}) 
\nn\\
&& =
\left[ 2 \sqrt{2} \frac{ \langle {\bf 2} | \ell_A | {\bf 2} \bigr]  }{M} \ell^{\mu}_{B} \epsilon_{\mu}(1) 
+ \epsilon(1) \cdot \epsilon({\bf 2})\Bigl( 2 \ell_A \cdot p_2 + M^2 \Bigr) \right]^2
\frac{i}{(\ell_A + p_2)^2 - m^2}
\nn\\
&& +  
\left[ 2 \sqrt{2} \frac{ \langle {\bf 2} | \ell_B | {\bf 2} \bigr]  }{M} \ell^{\mu}_{A} \epsilon_{\mu}(1)
+ \epsilon(1) \cdot \epsilon({\bf 2})  \Bigl( 2 \ell_A \cdot p_1 \Bigr) \right]^2
\frac{i}{(\ell_A + p_1)^2 - m^2} 
\nn\\
&& + \Bigl\{ 2 
\Bigl[  ( \epsilon(1) \cdot \epsilon({\bf 2}) )  ( p_1  - p_2 ) \cdot \ell_{A}
- 2 \Bigl( \ell_{A} \cdot \epsilon(1) p_1 \cdot \epsilon({\bf 2})
- \ell_{A} \cdot \epsilon({\bf 2}) p_2 \cdot \epsilon(1)  \Bigr) \Bigr]
- M^2 \epsilon(1) \cdot \epsilon({\bf 2}) \Bigr\}^2 
\frac{i}{(p_1+p_2)^2-M^2} .
\nn\\
\eea
The overall minus sign in the $t$-channel comes from the difference in signs of the associated numerators coming from the two different gauge theories in the double-copy map. Furthermore, similar to Eq.~(\ref{Mampscalar}), both expressions comprise all the Merlin degrees of freedom produced by the double copy flowing in the $t$-channel.

\section{Leading singularities in one-loop processes involving Merlin particles}

\begin{figure}[htb]
\begin{center}
\includegraphics[height=70mm,width=170mm]{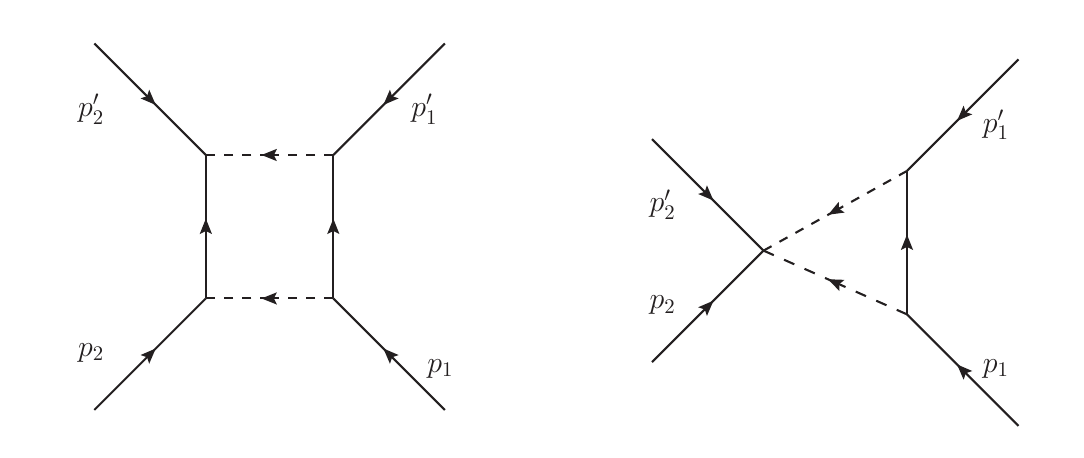}
\caption{Different topologies for the calculation of the leading singularities. Thick lines represent all possible external particles considered in the text, whereas dashed lines represent intermediate states.}
\label{LS}
\end{center}
\end{figure}

We are now in the position to calculate the leading singularities of one-loop amplitudes. As discussed above, we are interested in gluon as well as graviton scatterings in the corresponding higher-derivative theories. Moreover, we are also going to consider the one-loop $4$-point scattering process of massive scalars proceeding via exchange of gauge as well as gravitational particles. We will consider the narrow-width approximation for the associated Merlin propagators. We are going to consider two different topologies for the leading singularities, that is, box and triangle. These are depicted in Fig.~\ref{LS}.

Let us briefly outline the necessary steps to implement the technique before move on to the actual calculation. As discussed above, generalized unitarity explores discontinuities in loop diagrams. The important point is that, as in the case of standard unitarity cuts associated with two-particle exchanges, the additional cuts can also be realized as contour integrals. That this is a mandatory treatment for leading singularities can be noticed from the fact that unitarity methods requires the replacement of propagators by on-shell delta functions -- for more than two massless propagators, the solutions to the on-shell equations are complex, and the associated delta functions can only yield zero. So a well-defined procedure leads one to consider contour integrals instead of delta functions~\cite{Kosower:12}. Every time a residue is evaluated one uncovers a higher co-dimension singularity. The maximal number of residues at $L$-loop order is $4L$ in four dimensions. Evaluating all $4L$ residues generates the highest co-dimension singularity -- the corresponding discontinuity is the leading singularity~\cite{Cachazo:2008vp,Cachazo:2017jef}.

As mentioned above, the support of the delta functions is outside the physical region where the loop momentum is a real vector. This suggests that the integration procedure associated with the leading singularity should be carried out in terms of contour integrals in $\mathbb{C}^4$, and the loop momentum is therefore to be regarded as a complex vector. By doing the four-dimensional loop-momentum integral over each such contour one obtains the residue at the associated encircled pole. Contrary to the product of delta functions, the transformations here yield a factor of the inverse of the Jacobian. This ensures that we obtain an analytic result, and hence further contour integrations can be carried out~\cite{Cachazo:2008vp,Cachazo:2017jef}. As we will see in the course of the calculations, in our case there will be two distinct leading singularities.

As discussed in Ref.~\cite{Menezes:2021tsj}, unitarity-based methods are still useful in theories containing resonances; one just need to verify whether external momentum configurations of the amplitude allows the unstable propagator to become resonant. When one is off resonance, one possible way to deal with unstable particles is to eliminate them through the use of a description which contains only stable modes, albeit a non-local one~\cite{Menezes:2021tsj}. When the coupling to the decay products is very small, and therefore only resonance production is important, one may resort to the narrow-width approximation in order to circumvent such issues. 

On the other hand, in theories containing Merlin modes, close to the resonance the Merlin propagator has the generic form
$$
i D(q) \sim \frac{-i}{q^2 - m^2 - i \gamma}
$$
where $\gamma$ is associated with its decay width. As observed above, the presence of the two unusual minus signs is important for unitary -- this is a consequence of the fact that the imaginary part of the Merlin propagator is the same as of a normal resonance, as can be seen from the expression above:
$$
\textrm{Im}[ D(q) ] \sim \frac{- \gamma^2}{(q^2-m^2)^2 + \gamma^2}.
$$
Since we are interested in calculating contour integrals along adequate contours which are associated with the process of cutting propagators, which in turn yields the corresponding imaginary parts, two sets of amplitudes with different propagators which nevertheless have the same imaginary parts will 
generate identical results for the aforementioned contour integral. This means that the fact that leading singularities make sense in theories with unstable particles indicates that these are also available in the study of analytic properties of loop amplitudes containing Merlin modes. This was precisely one of the key points raised in Ref.~\cite{Menezes:2021tsj}.

A slight difference arises here. First, observe that, for stable particles and normal unstable particles, the sign of the residue of the one-particle pole is opposite in sign to the imaginary part of that pole. For the Merlin particle such signs are equal -- a similar result was also found in Ref.~\cite{Holdom:2021oii}. On the other hand, in the narrow-width approximation, the cut of a Merlin propagator should produce the same delta function as of a normal resonance. However, the Merlin propagator has an overall minus sign; when taking its propagator to define a variable $1/u$ and then integrate over a contour $|u| = \varepsilon$ encircling $u=0$ in the complex plane, the resulting residue has the opposite sign of the normal resonance. However, both operations should be equivalent. Such observations motivate us to define the contours for the leading singularities involving Merlins in such a way that their orientations should be the opposite to that of a normal resonance. This is necessary as positive energy flow associated with the Merlin propagates backward in time; furthermore, the residue at the Merlin pole is always negative and the sign of the width is always opposite from normal resonances. A related modification was put forward in Ref.~\cite{DM:19} when discussing unitarity in the framework of Veltman's largest time equation.

In a sense, the fact that amplitudes with ghost and normal resonances can be reconstructed by using unitarty-based methods can be seen as a non-trivial corollary of the results given in Refs.~\cite{DM:19,Veltman:63}. Indeed, from Veltman's work it is known that normal resonances satisfy unitarity to all orders in perturbation theory. Thence any discontinuity computed considering normal resonances in the intermediate states can be translated to a discontinuity with ghost resonances by using the fact that such resonances can be described in the same propagator with the coupling to the stable states described by the same self-energy, that is
$$
iD(q) = \frac{i}{q^2 - m^2 + \Sigma(q)-q^4/\Lambda^2}.
$$
Hence, it follows that, if the normal resonance satisfies the unitarity constraint, the ghost resonance must also. Therefore unitarity-based methods (including leading singularities) can be successfully applied to amplitudes involving all such resonances, as argued in Ref.~\cite{Menezes:2021tsj}.

\subsection{Higher-derivative Yang-Mills}

\subsubsection{Gluon scattering}

Here we will study a specific one-loop gluon scattering $g^+ g^+ \to g^+ g^+$ proceeding in the higher-derivative Yang-Mills theory. The leading singularity formula with gluons running in the loop is given by
\beq
\textrm{LS}^{(g)}_{g} = \sum_{h_1,h_3,h = \pm} 
\oint_{\Gamma} \frac{d^4 \ell}{(2\pi)^4 } \frac{1}{\ell^2(\ell+p_1^{\prime})^2 (\ell-p_1)^2}
A_{3}[1^{\prime +}, \ell^{-h}, - \ell_1^{h_1}] 
A_{3}[-\ell^{h}, 1^{+}, -\ell_3^{h_3}] 
A_{4}[2^{+},\ell_3^{-h_3}, \ell_1^{-h_1}, 2^{\prime +}]
\eeq
where $\ell_1 = \ell+p_1^{\prime}$ and $\ell_3 = p_1 - \ell$. For Merlins running in the loop we find that
\beq
\textrm{LS}^{(g)}_{M} =  
\oint_{-\Gamma} \frac{d^4 \ell}{(2\pi)^4} \frac{-1}{(\ell^2 - M^2)} 
\frac{-1}{[ (\ell+p_1^{\prime})^2 - M^2 ]}\frac{-1}{[ (\ell-p_1)^2 - M^2 ]}
A_{3}[1^{\prime +}, \boldsymbol\ell, - \boldsymbol\ell_1] 
A_{3}[-\boldsymbol\ell, 1^{+}, -\boldsymbol\ell_3] 
A_{4}[2^{+},\boldsymbol\ell_3, \boldsymbol\ell_1, 2^{\prime +}] .
\eeq
In the above formulae we take color-ordered amplitudes as the result with the full amplitudes can be obtained in the same way as will now be described. Notice also that these are the only possible configurations due to the fact that amplitudes with only one Merlin vanish. Generically the contour 
$\Gamma$ is defined as enclosing all the corresponding poles produced by cutting the propagators -- the minus in front of it in the expression for $\textrm{LS}^{(g)}_{M}$ is to remind us that the contour associated with a Merlin pole must have the opposite orientation to that corresponding to a stable particle (or a normal resonance).

We begin our discussion with the associated scattering with gluons running in the loop. First consider the triangle topology for the leading singularity. In principle, the only contribution in the sum over the helicities $h_1, h_3$ in $\textrm{LS}^{(g)}_{\triangle g}$ would be the term corresponding to $h_1 = h_3 = +$. However, a quick inspection of the remainder, that is
\beq
\textrm{LS}^{(g)}_{\triangle g} = \sum_{h = \pm} 
\oint_{\Gamma} \frac{d^4 \ell}{(2\pi)^4 } \frac{1}{\ell^2(\ell+p_1^{\prime})^2 (\ell-p_1)^2}
A_{3}[1^{\prime +}, \ell^{-h}, - \ell_1^{+}] 
A_{3}[-\ell^{h}, 1^{+}, -\ell_3^{+}] 
A_{4}[2^{+},\ell_3^{-}, \ell_1^{-}, 2^{\prime +}]
\eeq
reveals that it is actually zero. So in pure Yang-Mills theories the triangle leading singularity associated with the one-loop scattering $A^{1-\textrm{loop}}_{4}[1^{\prime +},1^{+},2^{+},2^{\prime +}]$ vanishes. What about the box leading singularity? With a judicious choice of reference spinors for the $3$-point amplitudes one can show that the associated product of $3$-point amplitudes vanishes. So the box leading singularity is also zero.

The fact that both leading singularities are zero should not come as a surprise. One-loop gluon amplitudes with gluons running in the loops can be calculated from the following identity~\cite{Britto:2010xq,Bern:1996ja,Brandhuber:05,Henn:14,Brandhuber:2006bf}
\beq
A^{1-\textrm{loop}} = A^{{\cal N} = 4} - 4 A^{{\cal N} = 1} + A^{\textrm{scalar}}
\eeq
where $A^{{\cal N} = 4}$ is the corresponding amplitude in ${\cal N} = 4$ super Yang-Mills theory, 
$A^{{\cal N} = 1}$ is the ${\cal N} = 1$ amplitude with a chiral multiplet in the loop and $A^{\textrm{scalar}}$ is the nonsupersymmetric amplitude with a complex scalar in the loop. Since all-plus gluon amplitude vanishes to all orders in perturbation theory for supersymmetric theories, $A^{1-\textrm{loop}}$ can be calculated by considering only complex scalars circulating in the loop -- but this amplitude consists of purely rational terms, which are cut-free in four dimensions. A non-trivial result can only be obtained by going to $D=4-2\epsilon$ dimensions. Hence its leading singularity must be zero in four dimensions.

Now let us study the case of Merlin particles running in the loop. We may proceed as in Ref.~\cite{Cachazo:2017jef} -- recall that the contour encircling a Merlin pole must have the opposite orientation as the one for a normal particle. We find that
\beq
\textrm{LS}^{(g)}_{M} =  - \frac{1}{8 ( p_1 \cdot p_1^{\prime} ) } 
\frac{1}{(2\pi i)}\oint_{\Gamma} \frac{dz}{z} 
A_{3}[1^{\prime +}, \boldsymbol\ell, - \boldsymbol\ell_1] 
A_{3}[-\boldsymbol\ell, 1^{+}, -\boldsymbol\ell_3] 
A_{4}[2^{+},\boldsymbol\ell_3, \boldsymbol\ell_1, 2^{\prime +}]  .
\eeq
The product of the amplitudes can be easily calculated:
\beq
A_{3}[1^{\prime +}, \boldsymbol\ell, - \boldsymbol\ell_1] 
A_{3}[-\boldsymbol\ell, 1^{+}, -\boldsymbol\ell_3] 
A_{4}[2^{+},\boldsymbol\ell_3, \boldsymbol\ell_1, 2^{\prime +}]
= - 24 M^{12}
\frac{\bigl[ 2 2^{\prime} \bigr]}{\langle 2 2^{\prime} \rangle}
\frac{\bigl[ 1 1^{\prime} \bigr]}{\langle 1 1^{\prime} \rangle} 
\frac{1}{(p_2+\ell_3)^2 - M^2}
\eeq
where we have factored out $g/M^2$. Observe that this expression corrects the overall numeric factor found in an analogous formula in Ref.~\cite{Menezes:2021dyp}. Using similar parameterizations as in Ref.~\cite{Cachazo:2017jef}, we find that
\beq
\textrm{LS}^{(g)}_{M} = M^{12}
\frac{\bigl[ 1 1^{\prime} \bigr]}{\langle 1 1^{\prime} \rangle} 
\frac{\bigl[ 2 2^{\prime} \bigr]}{\langle 2 2^{\prime} \rangle}
\frac{1}{(2\pi i)}\oint_{\Gamma} \frac{dz}{z} 
\frac{3 z}{ -2 (p_1 \cdot p_1^{\prime}) (\bar{q} \cdot p_2) z^2
+ 2 ( p_1 \cdot p_1^{\prime} ) ( p_1 \cdot p_2) z + M^2 q \cdot p_2 } .
\eeq
where $q^{\alpha \dot{\alpha}} = | 1^{\prime} \rangle \bigl[1|$ and the corresponding conjugate 
$\bar{q}^{\alpha \dot{\alpha}} = | 1 \rangle \bigl[1^{\prime}|$ are fixed reference massless vectors. We define the contour $\Gamma$ associated with the triangle topology as the contour enclosing either of the poles at $z=\infty$ and $z=0$. By choosing any of them will lead to $\textrm{LS}^{(g)}_{\triangle M} = 0$. On the other hand, by circling one of the two solutions of the quadratic factor that arises in the denominator will lead to the box topology. We obtain
\beq
\textrm{LS}^{(g)}_{\Box M} = - g^4
\frac{u t}{\langle 1 1^{\prime} \rangle \langle 1^{\prime} 2 \rangle \langle 2 2^{\prime} \rangle 
\langle 2^{\prime} 1\rangle}
\frac{6 M^{4}}{\sqrt{ s^2 u^2 + 4 M^2  s t u }}
\eeq
where the Mandelstam variables are defined as usual, that is, $s = (p_1+p_2)^2$, $t = (p_1 + p_2^{\prime})^2$ and $u = (p_1 + p_1^{\prime})^2$, with $s+t+u=0$. We also have restored the $g/M^2$ factors. So we see that the leading singularity associated with the one-loop scattering amplitude $A^{1-\textrm{loop}}_{4}[1^{\prime +},1^{+},2^{+},2^{\prime +}]$ does not vanish in the higher-derivative Yang-Mills theory. This is a consequence of the presence of massive Merlin modes running in the loops. Moreover, only the box topology is non-trivial -- essentially this is due to the fact that in the expression of the one-loop amplitude only the box integral is present~\cite{Menezes:2021dyp}.

\subsubsection{Scalar scattering}

Now we wish to explore general leading singularities in the scattering of matter particles $\phi \phi \to \phi \phi$ via gluon and Merlin exchange. For simplicity, we will work with identical massive scalars. The associated expressions are given by
\beq
\textrm{LS}^{(s)}_{g} = \sum_{h_1,h_3} 
\oint_{\Gamma} \frac{d^4 \ell}{(2\pi)^4 (\ell^2 - m^2)} \frac{1}{(\ell+p_1^{\prime})^2 (\ell-p_1)^2}
A_{3}({\bf 1}_j^{\prime}, \boldsymbol\ell, - \ell_1^{h_1}) 
A_{3}(-\boldsymbol\ell, {\bf 1}_i, -\ell_3^{h_3}) 
A_{4}({\bf 2}_m,\ell_3^{-h_3}, \ell_1^{-h_1}, {\bf 2}^{\prime}_n)
\eeq
for gluons running in the loop, whereas for Merlins we find that
\beq
\textrm{LS}^{(s)}_{M} =  
\oint_{-\Gamma} \frac{d^4 \ell}{(2\pi)^4 (\ell^2 - m^2)} 
\frac{-1}{[ (\ell+p_1^{\prime})^2 - M^2 ]}\frac{-1}{[ (\ell-p_1)^2 - M^2 ]}
A_{3}({\bf 1}_j^{\prime}, \boldsymbol\ell, - \boldsymbol\ell_1) 
A_{3}(-\boldsymbol\ell, {\bf 1}_i, -\boldsymbol\ell_3) 
A_{4}({\bf 2}_m,\boldsymbol\ell_3, \boldsymbol\ell_1, {\bf 2}_n^{\prime}) .
\eeq
where $m$ is the mass of the scalar particles. One also has two crossed terms:
\bea
\textrm{LS}^{(s)}_{gM} &=&  \sum_{h = \pm}
\oint_{-\Gamma} \frac{d^4 \ell}{(2\pi)^4 (\ell^2 - m^2)} 
\frac{1}{(\ell+p_1^{\prime})^2}\frac{-1}{[ (\ell-p_1)^2 - M^2 ]}
A_{3}({\bf 1}_j^{\prime}, \boldsymbol\ell, - \ell_1^{h}) 
A_{3}(-\boldsymbol\ell, {\bf 1}_i, -\boldsymbol\ell_3) 
A_{4}({\bf 2}_m,\boldsymbol\ell_3, \ell_1^{-h}, {\bf 2}_n^{\prime})
\nn\\
\textrm{LS}^{(s)}_{Mg} &=&  \sum_{h = \pm}
\oint_{-\Gamma} \frac{d^4 \ell}{(2\pi)^4 (\ell^2 - m^2)} 
\frac{-1}{[ (\ell+p_1^{\prime})^2 - M^2 ]} \frac{1}{(\ell-p_1)^2}
A_{3}({\bf 1}_j^{\prime}, \boldsymbol\ell, - \boldsymbol\ell_1) 
A_{3}(-\boldsymbol\ell, {\bf 1}_i, -\ell_3^{h}) 
A_{4}({\bf 2}_m,\ell_3^{-h}, \boldsymbol\ell_1, {\bf 2}_n^{\prime}) 
\eea
which contain a $4$-point amplitude with external Merlin and gluon, in addition to the scalars. There are actually other contributions that can be obtained from those above by considering $1 \leftrightarrow 2$. So the calculation of those above will suffice to our purposes.

We begin with $\textrm{LS}^{(s)}_{g}$. The technique is similar to the one used in Ref.~\cite{Cachazo:2017jef}. In the present case it will be instructive to describe it with some detail. We parametrize $\ell$ as
\beq
\ell = z L + \omega q 
\label{para}
\eeq
where $L^{\alpha \dot{\alpha}} = | L \rangle \bigl[L|$ and $q$ is a fixed reference massless momentum. Now the integration variables are $z,\omega \in \mathbb{C}$ and the helicity spinors $| L \rangle, | L \bigr]$. The measure becomes 
\bea
\frac{d^4 \ell}{\ell^2-m^2} = z dz \langle L dL \rangle \bigl[ L dL \bigr]
\frac{d\omega}{4 [ \omega-m^2/(2z L \cdot q ) ] } .
\eea
Contour integration around the pole $\ell^2=m^2$ fixes $\omega = m^2/(2z L \cdot q )$:
\beq
\textrm{LS}^{(s)}_{g} = \sum_{h_1,h_3} 
\frac{1}{4 (2\pi i)^3} \oint_{\Gamma} 
\frac{z dz \langle L dL \rangle \bigl[ L dL \bigr]}
{( 2 m^2 + 2 \ell \cdot p_1^{\prime} ) ( 2m^2 - 2 \ell \cdot p_1 )}
A_{3}({\bf 1}_j^{\prime}, \boldsymbol\ell, - \ell_1^{h_1}) 
A_{3}(-\boldsymbol\ell, {\bf 1}_i, -\ell_3^{h_3}) 
A_{4}({\bf 2}_m,\ell_3^{-h_3}, \ell_1^{-h_1}, {\bf 2}_n^{\prime}) .
\eeq
Now define two auxiliary massless vectors $p_3,p_4$ through the following relations
\bea
p_1^{\prime} &=& p_3 + x_1 p_4
\nn\\
p_1 &=& p_4 + x_1 p_3
\nn\\
x_1 &=& \frac{m^2}{2 p_3 \cdot p_4} .
\eea
From such definitions, one can show that
\bea
\frac{(1+x_1)^2}{x_1} &=& \frac{(p_1+p_1^{\prime})^2}{m^2}
\nn\\
\frac{(1-x_1)^2}{x_1} &=& \frac{(p_1+p_1^{\prime})^2-4m^2}{m^2} .
\eea
The reference massless vector $q$ and its conjugate $\bar{q}$ will be taken to be
\bea
q^{\alpha \dot{\alpha}} &=& | 3 \rangle \bigl[ 4|
\nn\\
\bar{q}^{\alpha \dot{\alpha}} &=& | 4 \rangle \bigl[ 3| .
\eea
Now use the expansion
\beq
L = A p_3 + B p_4 + C q + D \bar{q} .
\eeq
We choose $D=1$ as the overall scale of $L$ is immaterial, and we find that $C=AB$ as a consequence of $L^2 = 0$. One finds that
\beq
L^{\alpha \dot{\alpha}} = | L \rangle \bigl[L|
= \Bigl( A | 3 \rangle + |4\rangle \Bigr) \Bigl( \bigl[3| + B \bigl[4| \Bigr).
\eeq
Therefore, with a simple change of variables one obtains that
\beq
\textrm{LS}^{(s)}_{g} = \frac{x_1}{4 m^2} \frac{1}{(2\pi i)^3}
\sum_{h_1,h_3} \oint_{\Gamma} 
\frac{z dz dA dB }
{( 2 x_1 + z ( A x_1 + B ) ) ( 2 x_1 -  z ( A + B x_1 ) )}
A_{3}({\bf 1}_j^{\prime}, \boldsymbol\ell, - \ell_1^{h_1}) 
A_{3}(-\boldsymbol\ell, {\bf 1}_i, -\ell_3^{h_3}) 
A_{4}({\bf 2}_m,\ell_3^{-h_3}, \ell_1^{-h_1}, {\bf 2}_n^{\prime}) .
\eeq
Contour integrations over the two visible poles fix
$$
A = - B = \frac{2 x_1}{z(1-x_1)}
$$
and leads us to
\beq
\textrm{LS}^{(s)}_{g} = \frac{x_1}{4 m^2 (1-x_1^2)} 
\sum_{h_1,h_3} \frac{1}{(2\pi i)} \oint_{\Gamma} 
\frac{dz}{z}
A_{3}({\bf 1}_j^{\prime}, \boldsymbol\ell, - \ell_1^{h_1}) 
A_{3}(-\boldsymbol\ell, {\bf 1}_i, -\ell_3^{h_3}) 
A_{4}({\bf 2}_m,\ell_3^{-h_3}, \ell_1^{-h_1}, {\bf 2}_n^{\prime}) .
\eeq
Now let us explicitly evaluate the case with opposite helicities. Using the following parameterizations
\bea
\ell_1^{\alpha \dot{\alpha}}(z) &=& | \ell_1 \rangle \bigl[ \ell_1 |
= r(x_1) \left( | 3 \rangle + \frac{z}{r(x_1)} | 4 \rangle \right) 
\left( \bigl[ 3| - \frac{x_1}{z} r(x_1)  \bigl[ 4| \right)
\nn\\
\ell_3^{\alpha \dot{\alpha}}(z) &=& | \ell_3 \rangle \bigl[ \ell_3 |
= r(x_1) \left( | 4 \rangle + \frac{x_1}{z} r(x_1)  | 3 \rangle \right) 
\left( \bigl[ 4| - \frac{z}{r(x_1)}  \bigl[ 3| \right) 
\nn\\
r(x) &=& \frac{1+x}{1-x}
\eea
and the $3$-point amplitudes given above, one can prove that
\beq
\textrm{LS}^{(s+-)}_{g} = - \frac{g^2}{2} \frac{x_1^2}{\left(1-x_1^2\right)} 
\left( \frac{1+x_1}{1-x_1} \right)^2 T^{c}_{j k} T^{d}_{k i}
\frac{1}{2 \pi i} \oint_{\Gamma} \frac{dz}{z^3}
A_{4}({\bf 2}_m,\ell_3^{d +}, \ell_1^{c -}, {\bf 2}_n^{\prime})
\eeq
and
\beq
\textrm{LS}^{(s-+)}_{g} = - \frac{g^2}{2} \frac{1}{\left(1-x_1^2\right) r(x_1)^2}  T^{c}_{j k} T^{d}_{k i}
\frac{1}{2 \pi i} \oint_{\Gamma} dz \, z
A_{4}({\bf 2}_m,\ell_3^{d -}, \ell_1^{c +}, {\bf 2}_n^{\prime}) ,
\eeq
where the associated color factors coming from the $3$-point amplitudes are manifest. Let us focus on 
$\textrm{LS}^{(s+-)}_{g}$. For the purpose of calculating the leading singularity associated with the triangle topology with opposite helicities, it is useful to rewrite the Compton amplitude as
\bea
A_{4}({\bf 2}_{m},\ell_3^{d +}, \ell_1^{c -}, {\bf 2}_{n}^{\prime}) &=& 
= - \frac{2 g^2}{t}  \left( \frac{T^{c}_{ml} T^{d}_{ln}}{2 p_2 \cdot \ell_1} 
+ \frac{T^{d}_{ml} T^{c}_{ln}}{2 p_2^{\prime} \cdot \ell_1} \right)
\langle \ell_1 | {\bf 2} | \ell_3 \bigl]^{2}
\nn\\
&=&
- \frac{g^2}{t}  \left( \frac{ \{ T^{c} ,T^{d} \}_{mn} + [ T^{c} ,T^{d} ]_{mn} }{2 p_2 \cdot \ell_1} 
+ \frac{  \{ T^{d} ,T^{c} \}_{mn} + [ T^{d} ,T^{c} ]_{mn} }{2 p_2^{\prime} \cdot \ell_1} \right)
\langle \ell_1 | {\bf 2} | \ell_3 \bigl]^{2}
\nn\\
&=&
g^2 \{ T^{c} ,T^{d} \}_{mn}
z^2 \frac{( z^2/r(x_1)^2 \bar{k} \cdot p_2 + z/r(x_1) (p_3 - p_4) \cdot p_2 - k \cdot p_2 )^2}
{[ (\bar{k} \cdot p_2)/r(x_1) z^2 + (p_3 - x_1 p_4) \cdot p_2 z - x_1 r(x_1) k \cdot p_2 ] 
[ p_2 \leftrightarrow p_2^{\prime} ]}
\nn\\
&-& \frac{2 i g^2}{t} f^{cde} T^{e}_{mn} r(x_1) 
z \frac{( z^2/r(x_1)^2 \bar{k} \cdot p_2 + z/r(x_1) (p_3 - p_4) \cdot p_2 - k \cdot p_2 )^2}
{[ (\bar{k} \cdot p_2)/r(x_1) z^2 + (p_3 - x_1 p_4) \cdot p_2 z - x_1 r(x_1) k \cdot p_2 ] 
[ p_2 \leftrightarrow p_2^{\prime} ]}
\nn\\
&\times&
\left( z ( p_2' - p_2 ) \cdot ( p_3 - x_1 p_4 )
- x_1 r(x_1) ( p_2' - p_2 ) \cdot q
+ z^2/r(x_1)  ( p_2' - p_2 ) \cdot \bar{q} \right) 
\eea
where the Mandelstam variables are now given by
\bea
s &=& (p_1 + p_2)^2
\nn\\
t &=& (p_1 + p_1^{\prime})^2 = (p_2 + p_2^{\prime})^2
\nn\\
u &=& (p_1 + p_2^{\prime})^2 = (p_2 + p_1^{\prime})^2 
\label{mandelstam}
\eea
where
\beq
s + t + u = 4 m^2.
\eeq
Now define
$$
E^4 \equiv - 4 (1-x_1)^2 (q \cdot p_2) (\bar{q} \cdot p_2) = - s u .
$$
and introduce the change of variables 
$$
z = \frac{2 (1+x_1)}{E^2} \sqrt{-x_1} \, (q \cdot p_2) z^{\prime}
= - \frac{2 (1+x_1)}{E^2} \sqrt{-x_1} \, (q \cdot p_2^{\prime}) z^{\prime},
$$
to obtain that
\bea
\textrm{LS}^{(s+-)}_{g} &=& - g^4 \frac{m}{2 \sqrt{-t}} \frac{1}{\sqrt{4-t/m^2}} 
\left( \frac{N^2+2}{4 N^2} \delta_{mn}  \delta_{ij} 
+ \frac{(N^2-4)}{4N}  \delta_{mi} \delta_{jn}  \right)
\nn\\
&\times& \frac{1}{2 \pi i} \oint_{\Gamma} \frac{dz}{z}
\frac{\left( z^2 (-x_1)^{1/2} + \frac{s-u}{E^2}  z + \frac{1}{(-x_1)^{1/2}} \right)^2}
{ \left( 1 + \frac{\sqrt{-t}}{m}  \frac{u}{E^2} z - z^2  \right)
\left( 1 - \frac{\sqrt{-t}}{m}  \frac{s}{E^2} z - z^2  \right)}
\nn\\
&+& \frac{g^4}{4t}  \frac{E^2}{(4-t/m^2)^{3/2}} ( N \delta_{mi}\delta_{jn} - \delta_{ij} \delta_{mn} )
\nn\\
&\times& 
\frac{1}{2 \pi i} \oint_{\Gamma} \frac{dz}{z^2}
\frac{\left( z^2 (-x_1)^{1/2} + \frac{s-u}{E^2}  z + \frac{1}{(-x_1)^{1/2}} \right)^2
\left( 1 + \frac{\sqrt{-t}}{m}  \frac{(u-s)}{2E^2} z - z^2 \right)}
{ \left( 1 + \frac{\sqrt{-t}}{m}  \frac{u}{E^2} z - z^2  \right)
\left( 1 - \frac{\sqrt{-t}}{m}  \frac{s}{E^2} z - z^2  \right)}
\eea
where we used the following $SU(N)$ relation~\cite{Haber:2019sgz}
\beq
T^{c}_{ml} T^{c}_{j k} = \frac{1}{2} \left( \delta_{mk} \delta_{jl} - \frac{1}{N} \delta_{ml} \delta_{j k} \right).
\eeq
One can also use that~\cite{Haber:2019sgz}
\bea
T^c T^d &=& \frac{1}{2} \left[ \frac{1}{N} \delta^{cd} {\bf 1} + ( d^{cde} + i f^{cde} ) T^{e} \right]
\nn\\
f^{abc} f^{abd} &=& N \delta^{cd}
\nn\\
d^{abc} d^{abd} &=& \frac{N^2-4}{N} \,  \delta^{cd}
\eea
with $d^{cde}$ being the totally symmetric group invariant of $SU(N)$. 

Now let us discuss the choice of the contour $\Gamma$. The triangle topology is given by either of the contours that encircles the residues at $z=0,\infty$. For $\textrm{LS}^{(s+-)}_{g}$ we will choose 
$\Gamma = S^{1}_{\infty}$. On the other hand, with the replacement $z \to -1/z$, it is easy to see that we obtain the same integrand as the one for $\textrm{LS}^{(s+-)}_{g}$ but with the contour encircling $z=0$ and in the opposite direction. In other words, the conjugate contribution $\textrm{LS}^{(s-+)}_{g}$ is given by $\textrm{LS}^{(s+-)}_{g}$ with the replacement $S^{1}_{\infty} \to - S^{1}_{0}$. Finally, enclosing one of the poles produced by the quadratic factors in the denominator will produce the associated box topologies. 

The leading singularity associated with the triangle topology is then given by
\beq
\textrm{LS}^{(s)}_{\triangle g} = \textrm{LS}^{(s+-)}_{\triangle g} + \textrm{LS}^{(s-+)}_{\triangle g}
+ \textrm{LS}^{(s++)}_{\triangle g} + \textrm{LS}^{(s--)}_{\triangle g}
\eeq
where
\bea
\textrm{LS}^{(s+-)}_{\triangle g} + \textrm{LS}^{(s-+)}_{\triangle g} &=& 
- g^4 \frac{m}{2 \sqrt{-t}} \frac{1}{\sqrt{4-t/m^2}} 
\left( \frac{N^2+2}{4 N^2} \delta_{mn}  \delta_{ij} 
+ \frac{(N^2-4)}{4N}  \delta_{mi} \delta_{jn}  \right) \left( -2 + \frac{t}{m^2} \right)
\nn\\
&-& \frac{g^4}{8 m \sqrt{-t}} \frac{(s-u)}{(4-t/m^2)^{3/2}}  
( N \delta_{mi}\delta_{jn} - \delta_{ij} \delta_{mn} )
\left( -6 + \frac{t}{m^2} \right)
\label{LStri}
\eea
and the configurations with equal helicities will not contribute, since
\bea
\textrm{LS}^{(s++)}_{\triangle g} &=& \frac{x_1}{4 m^2 \left(1-x_1^2\right)} 
\frac{1}{2 \pi i} \oint_{(S^{1}_{\infty} - S^{1}_{0})} \frac{dz}{z}
A_{3}({\bf 1}_j^{\prime}, \boldsymbol\ell, - \ell_1^{+}) 
A_{3}(-\boldsymbol\ell, {\bf 1}_i, -\ell_3^{+}) 
A_{4}({\bf 2}_m,\ell_3^{-}, \ell_1^{-}, {\bf 2}_n^{\prime})
\nn\\
&=& \frac{g^4 m^2}{2}
\frac{x_1}{\left(1-x_1^2\right)} T^{c}_{j k} T^{d}_{k i}
\frac{1}{2 \pi i} \oint_{(S^{1}_{\infty} - S^{1}_{0})} dz
\left[\frac{T^{d}_{ml} T^{c}_{ln}}
{[ (\bar{k} \cdot p_2^{\prime}) z^2 + r(x_1) (p_3 - x_1 p_4) \cdot p_2^{\prime} z 
- x_1 r(x_1)^2 k \cdot p_2^{\prime} ] } 
\right.
\nn\\
&+& \left. \frac{T^{c}_{ml} T^{d}_{ln}}
{[ (\bar{k} \cdot p_2) z^2 + r(x_1) (p_3 - x_1 p_4) \cdot p_2 z - x_1 r(x_1)^2 k \cdot p_2 ] } \right]
\eea
and it is clear that this expression has zero residue at both $z=0,\infty$. A similar result is also valid for $h_1 = h_3=-1$. Observe the emergence of non-analytic factors in Eq.~(\ref{LStri}) -- the second term arises due to the non-Abelian nature of the interaction, which leads to the existence of a physical $t$-channel massless pole in the Compton amplitude involving gluons. The standard physical interpretation of the two branch points $t = 0$ and $t = 4m^2$ is that they correspond to the threshold for production of massless and massive states. This result agrees with the fact that triangle topology leading singularity is the double discontinuity across the t-channel~\cite{Cachazo:2017jef}.

As for the box topology, we write
\beq
\textrm{LS}^{(s+-)}_{\Box g} = \textrm{LS}^{(s+-)}_{1\Box g} + \textrm{LS}^{(s+-)}_{2\Box g}
\eeq
with the following definition
\beq
\textrm{LS}^{(s+-)}_{1\Box g} = \frac{2 g^4}{t} \frac{x_1^2}{\left(1-x_1^2\right)} r(x_1) 
\frac{1}{4} 
\left[ \delta_{ij} \delta_{mn} \left( 1 + \frac{1}{N^2} \right)
- \frac{2}{N} \delta_{mi} \delta_{jn} \right]
\frac{1}{2 \pi i} \oint_{S^{1}_{y_0}} \frac{dy}{y^2}
\frac{\left(  - \frac{E^4}{4 (1-x_1)^2} y^2  + (p_3 - p_4) \cdot p_2 \, y - 1 \right)^2}
{- \frac{E^4}{4 (1-x_1)^2} y^2 + (p_3 - x_1 p_4) \cdot p_2 \, y - x_1 }
\eeq
and $\textrm{LS}^{(s+-)}_{2\Box g}$ can be obtained from $\textrm{LS}^{(s+-)}_{1\Box g}$ by swapping 
$p_2 \leftrightarrow p_2^{\prime}$ and $c \leftrightarrow d$ in the Compton amplitude. In the above expression we performed the change of variables $z = r(x_1) (q \cdot p_2) y = - r(x_1) (q \cdot p_2^{\prime}) y$ and we used the definition of $E$. We also have used again the Fierz identity in $SU(N)$. By choosing $y_0$ to be one of the roots of the denominator, we finally find that
\bea
\textrm{LS}^{(s+-)}_{1\Box g} &=& \frac{g^4}{16} 
\left[ \left( 1 + \frac{1}{N^2} \right) \delta_{ij} \delta_{mn} 
- \frac{2}{N} \delta_{jn} \delta_{mi}  \right]
\frac{( u-2m^2+ \sqrt{E^4 - t u} )^2}{t \sqrt{E^4 - t u}} 
\nn\\
\textrm{LS}^{(s+-)}_{2\Box g} &=&
\frac{g^4}{16} 
\left[  \frac{1}{N^2}  \delta_{ij} \delta_{mn}  
+ \frac{N^2-2}{N} \delta_{jn} \delta_{mi}
 \right]
\frac{( s-2m^2+ \sqrt{E^4 - t s} )^2}{t \sqrt{E^4 - t s}} .
\eea
One can prove that the reverse helicity configuration can be obtained from the previous expression by taking the other root of the denominator; therefore, with a similar decomposition we find that
\bea
\textrm{LS}^{(s-+)}_{1\Box g} &=& \frac{g^4}{16} 
\left[ \left( 1 + \frac{1}{N^2} \right) \delta_{ij} \delta_{mn} 
- \frac{2}{N} \delta_{jn} \delta_{mi}  \right]
\frac{( u - 2 m^2 - \sqrt{E^4 - t u})^2}{t \sqrt{E^4 - t u}}
\nn\\
\textrm{LS}^{(s-+)}_{2\Box g} &=&
\frac{g^4}{16} 
\left[  \frac{1}{N^2}  \delta_{ij} \delta_{mn}  
+ \frac{N^2-2}{N} \delta_{jn} \delta_{mi}
 \right]
\frac{( s - 2 m^2 - \sqrt{E^4 - t s})^2}{t \sqrt{E^4 - t s}} .
\eea
Analogous decompositions and calculations for configurations with equal helicity yield
\bea
\textrm{LS}^{(s++)}_{1\Box g} &=& \frac{g^4}{4} 
\left[ \left( 1 + \frac{1}{N^2} \right) \delta_{ij} \delta_{mn} 
- \frac{2}{N} \delta_{jn} \delta_{mi}  \right]
\frac{m^4}{t \sqrt{E^4 - t u}}
\nn\\
\textrm{LS}^{(s++)}_{2\Box g} &=&
\frac{g^4}{4} 
\left[  \frac{1}{N^2}  \delta_{ij} \delta_{mn}  
+ \frac{N^2-2}{N} \delta_{jn} \delta_{mi}
 \right]
\frac{m^4}{t \sqrt{E^4 - t s}} .
\eea
It is easy to see that the minus configuration produces the same result. Notice that the box topology yields a pole $1/t$, which would enable one to extend the computation to an $r$-loop ladder, just as the gravitational case~\cite{Cachazo:2017jef}. Such outcomes are in agreement with the fact that the box topology leading singularity can be envisaged as the discontinuity in the $t$-channel of the function derived from the calculation of the discontinuity in the $s$-channel of the one-loop amplitude~\cite{Cachazo:2017jef}.

Now let us calculate the one-loop leading singularity with only Merlins in the loop. By using the same method as before, we find that
\bea
\textrm{LS}^{(s)}_{M} &=&  \frac{x_1}{4 m^2} 
\frac{1}{(2\pi i)^3} \oint_{-\Gamma} z dz dA dB 
\frac{-1}{ \bigl( (2 - M^2/m^2) x_1 + z ( A x_1 + B ) \bigr)} 
\frac{-1}{\bigl( (2 - M^2/m^2) x_1 - z ( A + B x_1 ) \bigr)}
\nn\\
&\times& A_{3}({\bf 1}_j^{\prime}, \boldsymbol\ell, - \boldsymbol\ell_1) 
A_{3}(-\boldsymbol\ell, {\bf 1}_i, -\boldsymbol\ell_3) 
A_{4}({\bf 2}_m,\boldsymbol\ell_3, \boldsymbol\ell_1, {\bf 2}_n^{\prime}) 
\eea
so that the contour integrations in the $A,B$ planes should be carried out along contours with the opposite direction as the ones taken in the gluon case. Such contour integrations over $A,B$ fix
\beq
A = - B = \frac{ ( 2 - M^2/m^2 ) x_1}{ z(1-x_1) }
\eeq
and yields
\beq
\textrm{LS}^{(s)}_{M} =  \frac{x_1}{4 m^2 (1-x_1^2)} 
\frac{1}{(2\pi i)} \oint_{\Gamma} 
\frac{dz} {z}
A_{3}({\bf 1}_j^{\prime}, \boldsymbol\ell, - \boldsymbol\ell_1) 
A_{3}(-\boldsymbol\ell, {\bf 1}_i, -\boldsymbol\ell_3) 
A_{4}({\bf 2}_m,\boldsymbol\ell_3, \boldsymbol\ell_1, {\bf 2}_n^{\prime}) .
\eeq
The product of the amplitudes can be evaluated by using the results listed above. We choose 
$\Gamma = S^{1}_{\infty}$ to evaluate the triangle leading singularity. We find that
\bea
\textrm{LS}^{(s)}_{\triangle M} &=& - \frac{g^4}{16 m} \frac{1}{\sqrt{-t}} \frac{1}{\sqrt{4-t/m^2}} 
\Biggl[ \left( \frac{N^2-2}{N} \delta_{jn}  \delta_{mi} + \frac{1}{N^2} \delta_{ij} \delta_{mn} \right)
\frac{ 2 M^2 u + 2 s^2 -s t - 2 u (t+u) }{(s+u)}
\nn\\
&+& \left(  \frac{N^2+1}{N^2} \delta_{ij} \delta_{mn}  - \frac{2}{N}  \delta_{jn} \delta_{mi} \right)
\frac{ 2 M^2 s + 2 u^2 - u t - 2 s (t+s) }{(s+u)}
\nn\\
&+&( N \delta_{mi} \delta_{jn}  -  \delta_{ij} \delta_{mn} ) 
\frac{\left(2 M^2+t\right) (s-u) \left(-2 M^2+2 s+t+2 u\right)}{2 (s+u)}
\left( \frac{ 1 }{t} - \frac{ 2 }{t - M^2} \right) 
\nn\\
&-& \frac{1}{2} \left( \frac{N^2-4}{N} \delta_{mi} \delta_{jn}  +  
\frac{N^2+2}{N^2} \delta_{ij} \delta_{mn} \right) (t-8 m^2+2 M^2)   \Biggr]
\eea
where we have used $SU(N)$ identities given above and the results
\bea
(p_3 - x_1 p_4) \cdot p_2 &=&
\frac{r(x_1)}{2} u
\nn\\
(p_3 - x_1 p_4) \cdot p_2^{\prime} &=&
\frac{r(x_1)}{2} s
\nn\\
p_4 - p_3 &=& \frac{p_1 - p_1^{\prime}}{(1 - x_1)} .
\eea
Observe the emergence again of non-analytic factors; but there is an important difference here: We verify the presence of the pole at $t=M^2$ due to the presence of the Merlin modes in the $t$-channel of the Compton amplitude.

As discussed above, the box topology is obtained by considering the poles associated with the roots of the quadratic factors in the denominators. In this case again we use the definition of $E$. We obtain
\bea
\textrm{LS}^{(s)}_{\Box M} &=& - \frac{g^4}{16} \frac{1}{\sqrt{-t}}
\left[ \left( \frac{N^2-2}{N} \delta_{jn}  \delta_{mi} + \frac{1}{N^2} \delta_{ij} \delta_{mn} \right)
\frac{  \left(4 m^2-M^2-2 s\right)^2}{\sqrt{ \left(E^4 - s t \right) \left(u-(t+u) \left(1-\frac{2 M^2}{t+u}\right)^2\right) } }
\right.
\nn\\
&+& \left.
\left(  \frac{N^2+1}{N^2} \delta_{ij} \delta_{mn}  - \frac{2}{N}  \delta_{jn} \delta_{mi} \right)
\frac{ \left(4 m^2-M^2-2 u\right)^2}{\sqrt{ \left(E^4 - u t\right) \left(s-(t+s) \left(1-\frac{2 M^2}{t+s}\right)^2\right) } }
\right] .
\eea
One interesting thing to observe is that the terms associated with the $t$-channel and the contact term of the Compton amplitude do not contribute to the box leading singularity, only to the triangle one. On the other hand, notice the appearance of the non-analytic factor $\sqrt{-t}$; this is not present in the box topology involving only gluons (which instead develops a pole $1/t$) and is clearly due to the presence of the Merlin modes. So the Merlin particle modifies in a non-trivial way the analytic structure of the amplitude, the crucial difference being the presence of an additional branch cut.

Now let us calculate the crossed terms. Using the same technique, we find that
\bea
\textrm{LS}^{(s)}_{gM} &=&  \sum_{h = \pm}
\frac{x_1}{4 m^2} 
\frac{1}{(2\pi i)^3} \oint_{-\Gamma} z dz dA dB 
\frac{1}{ \bigl( 2 x_1 + z ( A x_1 + B ) \bigr)} 
\frac{-1}{\bigl( (2 - M^2/m^2) x_1 - z ( A + B x_1 ) \bigr)}
\nn\\
&\times& A_{3}({\bf 1}_j^{\prime}, \boldsymbol\ell, - \ell_1^{h}) 
A_{3}(-\boldsymbol\ell, {\bf 1}_i, -\boldsymbol\ell_3) 
A_{4}({\bf 2}_m,\boldsymbol\ell_3, \ell_1^{-h}, {\bf 2}_n^{\prime})
\eea
and
\bea
\textrm{LS}^{(s)}_{Mg} &=&  \sum_{h = \pm}
\frac{x_1}{4 m^2} 
\frac{1}{(2\pi i)^3} \oint_{-\Gamma} z dz dA dB 
\frac{-1}{ \bigl( (2 - M^2/m^2) x_1 + z ( A x_1 + B ) \bigr)} 
\frac{1}{\bigl( 2 x_1 - z ( A + B x_1 ) \bigr)}
\nn\\
&\times& A_{3}({\bf 1}_j^{\prime}, \boldsymbol\ell, - \boldsymbol\ell_1) 
A_{3}(-\boldsymbol\ell, {\bf 1}_i, -\ell_3^{h}) 
A_{4}({\bf 2}_m,\ell_3^{-h}, \boldsymbol\ell_1, {\bf 2}_n^{\prime}) .
\eea
Carrying out the contour integrations over A and B (taking care to consider proper orientations for the contours as discussed above) yields
\beq
\textrm{LS}^{(s)}_{gM} =  - \frac{x_1}{4 m^2 \left(1-x_1^2\right) } 
\sum_{h = \pm}
\frac{1}{(2\pi i)} \oint_{\Gamma} 
\frac{dz}{ z }
A_{3}({\bf 1}_j^{\prime}, \boldsymbol\ell, - \ell_1^{h}) 
A_{3}(-\boldsymbol\ell, {\bf 1}_i, -\boldsymbol\ell_3) 
A_{4}({\bf 2}_m,\boldsymbol\ell_3, \ell_1^{-h}, {\bf 2}_n^{\prime})
\eeq
with
\bea
A_{gM} &=& \frac{x_1 \left(M^2-2 m^2 (x_1+1)\right)}{m^2 \left(x_1^2-1\right) z}
\nn\\
B_{gM} &=& - \frac{x_1 \left(M^2 x_1 - 2 m^2 (x_1+1)\right)}{m^2 \left(x_1^2-1\right) z}
\label{96}
\eea
and
\beq
\textrm{LS}^{(s)}_{Mg} = \frac{x_1}{4 m^2 \left(1-x_1^2\right)} 
\sum_{h = \pm}
\frac{1}{(2\pi i)} \oint_{\Gamma} 
\frac{dz}{z}
A_{3}({\bf 1}_j^{\prime}, \boldsymbol\ell, - \boldsymbol\ell_1) 
A_{3}(-\boldsymbol\ell, {\bf 1}_i, -\ell_3^{h}) 
A_{4}({\bf 2}_m,\ell_3^{-h}, \boldsymbol\ell_1, {\bf 2}_n^{\prime})
\eeq
with
\bea
A_{Mg} &=& \frac{x_1 \left(M^2 x_1 - 2 m^2 (x_1+1)\right)}{m^2 \left(x_1^2-1\right) z}
= - B_{gM} 
\nn\\
B_{Mg} &=& - \frac{x_1 \left(M^2-2 m^2 (x_1+1)\right)}{m^2 \left(x_1^2-1\right) z} 
= - A_{gM}.
\label{98}
\eea
The results presented above allows us to easily calculate the associated product of amplitudes that appear in our expression for the leading singularity associated with the crossed terms. After tedious algebraic manipulations, and again choosing $\Gamma = S^{1}_{\infty}$ to evaluate the triangle leading singularity, we find that
\bea
\textrm{LS}^{(s)}_{\triangle gM} &=& \frac{g^4}{8 m} \frac{1}{\sqrt{-t}} \frac{1}{\sqrt{4-t/m^2}}
\Biggl[  \left( \frac{N^2-2}{N} \delta_{jn}  \delta_{mi} + \frac{1}{N^2} \delta_{ij} \delta_{mn} \right)
 \left(s-2 m^2\right)
+  \left(  \frac{N^2+1}{N^2} \delta_{ij} \delta_{mn}  - \frac{2}{N}  \delta_{jn} \delta_{mi} \right)
 \left(u-2 m^2\right)
\nn\\
&-& ( N \delta_{mi} \delta_{jn} - \delta_{mn} \delta_{ji} )
\biggl( \frac{M^4 (u-s)+M^2 (s+u) (s-t-u)+t (s-u) (2 s+t+2 u)}{2 (s+u)}
\biggr) \frac{1}{t-M^2}
\nn\\
&-& \frac{1}{2}  \left( \frac{N^2-4}{N} \delta_{mi} \delta_{jn} + \frac{N^2+2}{N^2} \delta_{ij} \delta_{mn} \right)
\left( t-4 m^2+M^2 \right)
\Biggr]
\eea
and
\bea
\textrm{LS}^{(s)}_{\triangle Mg} &=& - \frac{g^4}{8 m} \frac{1}{\sqrt{-t}} \frac{1}{\sqrt{4-t/m^2}}
\Biggl[  \left(  \frac{N^2+1}{N^2} \delta_{ij} \delta_{mn}  - \frac{2}{N}  \delta_{jn} \delta_{mi} \right)
\left(u-2 m^2\right)
+  \left( \frac{N^2-2}{N} \delta_{jn}  \delta_{mi} + \frac{1}{N^2} \delta_{ij} \delta_{mn} \right)
 \left(s-2 m^2\right)
\nn\\
&-& ( N \delta_{mi} \delta_{jn} - \delta_{mn} \delta_{ji} )
\biggl( \frac{M^2 \left(s^2+u (2 t-u)\right)+t (s-u) (2 s+t+2 u)}{2 (s+u)} \biggr) \frac{1}{t-M^2}
\nn\\
&-& \frac{1}{2}  \left( \frac{N^2-4}{N} \delta_{mi} \delta_{jn}  +  
\frac{N^2+2}{N^2} \delta_{ij} \delta_{mn} \right) 
 \left(t-4 m^2\right) 
\Biggr]
\eea
where we used that
$$
(p_4 - p_3 x_1) \cdot p_2^{\prime} = \frac{r(x_1)}{2} u .
$$
Again observe the appearance of a pole at $t=M^2$.

Now, let us choose a contour encircling one of the poles associated with the roots of the quadratic factors in the denominators. The non-vanishing contribution reads
\bea
\textrm{LS}^{(s)}_{\Box g M} &=&  \frac{g^4}{4} \frac{\sqrt{(4m^2-t)^2}}{\sqrt{4m^2-t}}
\Biggl[ \left( \frac{N^2-2}{N} \delta_{jn}  \delta_{mi} + \frac{1}{N^2} \delta_{ij} \delta_{mn} \right)
\frac{\left(s-2 m^2\right)^2}{\sqrt{m^2 s \left(M^2-t\right)^2 (-s t-4 u)} }
\nn\\
&+& \left(  \frac{N^2+1}{N^2} \delta_{ij} \delta_{mn}  - \frac{2}{N}  \delta_{jn} \delta_{mi} \right)
\frac{\left(u-2 m^2\right)^2}{\sqrt{ m^2 u \left(M^2-t\right)^2 (-u t-4 s)} }
\Biggr]
\eea
and
\bea
\textrm{LS}^{(s)}_{\Box Mg} &=& - \frac{g^4}{4} \frac{\sqrt{(4m^2-t)^2}}{\sqrt{4m^2-t}}
\Biggl[  \left(  \frac{N^2+1}{N^2} \delta_{ij} \delta_{mn}  - \frac{2}{N}  \delta_{jn} \delta_{mi} \right)
\frac{\left(u-2 m^2\right)^2}{\sqrt{m^2 u \left(M^2-t\right)^2 (-4 s-t u)}}
\nn\\
&+&
 \left( \frac{N^2-2}{N} \delta_{jn}  \delta_{mi} + \frac{1}{N^2} \delta_{ij} \delta_{mn} \right)
\frac{\left(s-2 m^2\right)^2}{\sqrt{m^2 s \left(M^2-t\right)^2 (-4 u-t s)}}
\Biggr] .
\eea
Observe that in principle the crossed terms for the box topology also presents non-analytic terms instead of the standard pole $1/t$; however, it is trivial to see that the sum of both contributions vanish, which implies that only the triangle leading singularity is important for such crossed terms -- even for the triangle topology we observe non-trivial cancelations when we sum the two terms above.

\subsection{Quadratic gravity}

\subsubsection{Graviton scattering}

Now let us consider the associated gravity leading singularities. We consider a specific one-loop graviton-graviton scattering $h^{++} h^{++} \to h^{++} h^{++}$. With gravitons running in the loop, we find that the leading singularity is given by
\beq
\textrm{LS}^{(h)}_{h} = \sum_{h_1,h_3,h = ++,--} 
\oint_{\Gamma} \frac{d^4 \ell}{(2\pi)^4 } \frac{1}{\ell^2(\ell+p_1^{\prime})^2 (\ell-p_1)^2}
M_{3}(1^{\prime ++}, \ell^{-h}, - \ell_1^{h_1}) 
M_{3}(-\ell^{h}, 1^{++}, -\ell_3^{h_3}) 
M_{4}(2^{++},\ell_3^{-h_3}, \ell_1^{-h_1}, 2^{\prime ++}) .
\eeq
For Merlins running in the loop we find that
\beq
\textrm{LS}^{(h)}_{M} =  
\oint_{-\Gamma} \frac{d^4 \ell}{(2\pi)^4} \frac{-1}{(\ell^2 - M^2)} 
\frac{-1}{[ (\ell+p_1^{\prime})^2 - M^2 ]}\frac{-1}{ [ (\ell-p_1)^2 - M^2 ]}
M_{3}(1^{\prime ++}, \boldsymbol\ell, - \boldsymbol\ell_1)
M_{3}(-\boldsymbol\ell, 1^{++}, -\boldsymbol\ell_3) 
M_{4}(2^{++},\boldsymbol\ell_3, \boldsymbol\ell_1, 2^{\prime ++}) .
\eeq
The double-copy prescription allows us to recycle the gluon results derived previously and we find that 
$\textrm{LS}^{(h)}_{h} = 0$. Essentially the reason is the same as before: A similar supersymmetric decomposition allows one to see that the the one-loop scattering amplitude $M^{1-\textrm{loop}}_{4}[1^{\prime ++},1^{++},2^{++},2^{\prime ++}]$ consists of purely rational terms, which are cut-free in four dimensions~\cite{Dunbar:1994bn}. Hence the associated leading singularity must be zero in four dimensions. 

For the case of Merlins running in the loop, we will need the double-copy results presented above, as well as taking the adequate orientation for the associated contours; proceeding as in the analogous gauge-theory case, we find that
\bea
\textrm{LS}^{(h)}_{M} &=&  - \frac{1}{8 ( p_1 \cdot p_1^{\prime} ) } 
\frac{1}{(2\pi i)}\oint_{\Gamma} \frac{dz}{z} 
M_{3}(1^{\prime ++}, \boldsymbol\ell, - \boldsymbol\ell_1) 
M_{3}(-\boldsymbol\ell, 1^{++}, -\boldsymbol\ell_3) 
M_{4}(2^{++},\boldsymbol\ell_3, \boldsymbol\ell_1, 2^{\prime ++})
\nn\\
&=& - \left( \delta^{(I}_{I} \delta^{J}_{J} \delta^{K}_{K} \delta^{L)}_{L} \right) 
\frac{2 i M^{8}}{ ( p_1 \cdot p_1^{\prime} ) } 
\frac{\bigl[ 2 2^{\prime} \bigr]^2}{\langle 2 2^{\prime} \rangle^2}
\frac{\bigl[11^{\prime}\bigr]^2}{\langle 1 1^{\prime} \rangle^2} 
\frac{1}{(2\pi i)}\oint_{\Gamma} \frac{dz}{z} 
\left( \frac{1}{2 p_2^{\prime} \cdot \ell_1} + \frac{1}{2 p_2 \cdot \ell_1} \right)
\nn\\
&=& i M^{8} \left( \delta^{(I}_{I} \delta^{J}_{J} \delta^{K}_{K} \delta^{L)}_{L} \right)
\frac{\bigl[ 2 2^{\prime} \bigr]^2}{\langle 2 2^{\prime} \rangle^2}
\frac{\bigl[11^{\prime}\bigr]^2}{\langle 1 1^{\prime} \rangle^2} 
\frac{1}{(2\pi i)}\oint_{\Gamma} \frac{dz}{z} 
\left( \frac{2 z}{ - 2 ( p_1 \cdot p_1^{\prime} ) (\bar{q} \cdot p_2^{\prime}) z^2
- 2 ( p_1 \cdot p_1^{\prime} ) (p_1^{\prime}\cdot p_2^{\prime}) z
+ M^2 q \cdot p_2^{\prime}} 
\right.
\nn\\
&+& \left. 
\frac{2 z}{ - 2 ( p_1 \cdot p_1^{\prime} ) (\bar{q} \cdot p_2) z^2 
- 2 ( p_1 \cdot p_1^{\prime} ) (p_1^{\prime}\cdot p_2) z
+ M^2 q \cdot p_2} \right) .
\eea
In Ref.~\cite{Menezes:2021dyp} a similar product of amplitudes was considered, except that the result in there lacks the above overall factor given by a (non-normalized) symmetric product of the delta functions, so this expression slightly corrects the analogous one in such a reference. As in previous case, we see that $\textrm{LS}^{(h)}_{\triangle M} = 0$. The box topology can be easily calculated by relying on the results derived above:
\beq
\textrm{LS}^{(h)}_{\Box M} = \left( \delta^{(I}_{I} \delta^{J}_{J} \delta^{K}_{K} \delta^{L)}_{L} \right)
\frac{i \kappa^4 M^{8}}{4} 
\left( \frac{u t}{\langle 1 1^{\prime} \rangle \langle 1^{\prime} 2 \rangle \langle 2 2^{\prime} \rangle 
\langle 2^{\prime} 1\rangle} \right)^2
\left( \frac{1}{\sqrt{ s^2 u^2 + 4 M^2  s t u }}
+ \frac{1}{\sqrt{ t^2 u^2 + 4 M^2  s t u }}
\right) 
\eeq
where the Mandelstam variables are defined as $s = (p_1+p_2)^2$, 
$t = (p_1 + p_2^{\prime})^2$ and $u = (p_1 + p_1^{\prime})^2$, with $s+t+u=0$. We also have restored the factors $\kappa/2$ coming from the amplitudes. As in the previous case, in the one-loop amplitude only the box integral is present~\cite{Menezes:2021dyp} -- as a result, only the box topology is non-vanishing.

\subsubsection{Scalar scattering}

To conclude our exploration, let us now study leading singularities in the scattering of identical scalar particles $\phi \phi \to \phi \phi$ interacting gravitationally. The associated expressions are given by
\beq
\textrm{LS}^{(s)}_{h} = \sum_{h_1,h_3 = ++,--} 
\oint_{\Gamma} \frac{d^4 \ell}{(2\pi)^4 (\ell^2 - m^2)} \frac{1}{(\ell+p_1^{\prime})^2 (\ell-p_1)^2}
M_{3}({\bf 1}^{\prime}, \boldsymbol\ell, - \ell_1^{h_1}) 
M_{3}(-\boldsymbol\ell, {\bf 1}, -\ell_3^{h_3}) 
M_{4}({\bf 2},\ell_3^{-h_3}, \ell_1^{-h_1}, {\bf 2}^{\prime})
\eeq
for gravitons running in the loop, whereas for Merlins we find that
\beq
\textrm{LS}^{(s)}_{M} =  
\oint_{-\Gamma} \frac{d^4 \ell}{(2\pi)^4 (\ell^2 - m^2)} 
\frac{-1}{[ (\ell+p_1^{\prime})^2 - M^2 ]}\frac{-1}{ [ (\ell-p_1)^2 - M^2 ]}
M_{3}({\bf 1}^{\prime}, \boldsymbol\ell, - \boldsymbol\ell_1) 
M_{3}(-\boldsymbol\ell, {\bf 1}, -\boldsymbol\ell_3) 
M_{4}({\bf 2},\boldsymbol\ell_3, \boldsymbol\ell_1, {\bf 2}^{\prime}) 
\eeq
where $m$ is the mass of the scalar particles. We consider all possible Merlins here -- the ones associated with the graviton, the dilaton and the Kalb-Ramond fields. This amounts to take the total Merlin propagator in the form
\begin{equation}
D_{\mu\rho; \nu\sigma}(p) \bigg|_{\textrm{Merlin}} = 
- \frac{1}{(p^2 - M^2)}  \widetilde{\pi}_{\mu\rho} \widetilde{\pi}_{\nu\sigma} 
\end{equation}
which would only contract ``left" indices with ``right" indices. $3$-point vertices would also have to be constructed in this way, as a product of $3$-point (higher-derivative) Yang-Mills vertices. Such a factorization would make the double-copy property manifest. Similar choices for the graviton propagator  and the associated $3$-point vertex were proposed in Ref.~\cite{Bern:1999ji}.

As above, one also has to consider two crossed terms:
\bea
\textrm{LS}^{(s)}_{hM} &=&  \sum_{h}
\oint_{-\Gamma} \frac{d^4 \ell}{(2\pi)^4 (\ell^2 - m^2)} 
\frac{1}{(\ell+p_1^{\prime})^2}\frac{-1}{ [ (\ell-p_1)^2 - M^2 ]}
M_{3}({\bf 1}^{\prime}, \boldsymbol\ell, - \ell_1^{h}) 
M_{3}(-\boldsymbol\ell, {\bf 1}, -\boldsymbol\ell_3) 
M_{4}({\bf 2},\boldsymbol\ell_3, \ell_1^{-h}, {\bf 2}^{\prime})
\nn\\
\textrm{LS}^{(s)}_{Mh} &=&  \sum_{h}
\oint_{-\Gamma} \frac{d^4 \ell}{(2\pi)^4 (\ell^2 - m^2)} 
\frac{-1}{[ (\ell+p_1^{\prime})^2 - M^2 ]}\frac{1}{ (\ell-p_1)^2}
M_{3}({\bf 1}^{\prime}, \boldsymbol\ell, - \boldsymbol\ell_1) 
M_{3}(-\boldsymbol\ell, {\bf 1}, -\ell_3^{h}) 
M_{4}({\bf 2},\ell_3^{-h}, \boldsymbol\ell_1, {\bf 2}^{\prime}) 
\eea
which contain a $4$-point amplitude with external Merlin and graviton, in addition to the scalars. 
Observe that in such crossed terms all degrees of freedom generated by the double-copy procedure are being taken into account. Concerning the graviton contribution, this would be equivalent to taking its propagator in the double-copy form~\cite{Bern:1999ji}
\begin{equation}
D_{\mu\rho; \nu\sigma}(p) \bigg|_{\textrm{graviton}} = 
 \frac{1}{p^2}  \eta_{\mu\rho} \eta_{\nu\sigma} 
\end{equation}
and the terminology ``graviton" above could apply to all massless degrees of freedom produced by the double copy -- unless we impose that all vertices in the $n$-graviton scattering amplitude should satisfy a rigid left-right interchange symmetry, as explained in Ref.~\cite{Bern:1999ji}. As in the gauge-theory case, there are also other contributions that can be obtained from those above by considering $1 \leftrightarrow 2$.

Let us begin with the case of only gravitons running in the loop. Since this was already calculated before, we simply quote the result~\cite{Cachazo:2017jef}. For the triangle topology, we find that
\beq
\textrm{LS}^{(s)}_{\triangle h} = \textrm{LS}^{(s ++,--)}_{\triangle h} + \textrm{LS}^{(s --,++)}_{\triangle h}
+ \textrm{LS}^{(s ++,++)}_{\triangle h} + \textrm{LS}^{(s --,--)}_{\triangle h}
\eeq
where
\beq
\textrm{LS}^{(s ++,--)}_{\triangle h} + \textrm{LS}^{(s --,++)}_{\triangle h} = 
i \left( \frac{\kappa}{2} \right)^4 \frac{m}{\sqrt{-t}} \frac{m^4}{(4-t/m^2)^{5/2}} 
\frac{1}{2 \pi i} \oint_{(S^{1}_{\infty} - S^{1}_{0})} \frac{dz}{z^3}
\frac{\left( z^2 (-x_1)^{1/2} + \frac{s-u}{E^2}  z + \frac{1}{(-x_1)^{1/2}} \right)^4}
{ \left( 1 + \frac{\sqrt{-t}}{m}  \frac{u}{E^2} z - z^2  \right)
\left( 1 - \frac{\sqrt{-t}}{m}  \frac{s}{E^2} z - z^2  \right)}
\eeq
and the contributions with equal helicities vanish. The box topology can be calculated in much the same way as was done in the Yang-Mills case; the final result reads
\beq
\textrm{LS}^{(s)}_{\Box h} = \textrm{LS}^{(s ++,--)}_{\Box h} + \textrm{LS}^{(s --,++)}_{\Box h}
+ \textrm{LS}^{(s ++,++)}_{\Box h} + \textrm{LS}^{(s --,--)}_{\Box h}
\eeq
with
\bea
\textrm{LS}^{(s ++,--)}_{\Box h} &=& i G^2 \pi^2 
\frac{( s-2m^2 + \sqrt{E^4 - t s} )^4}{t \sqrt{E^4 - t s}}
\nn\\
\textrm{LS}^{(s --,++)}_{\Box h} &=& G^2 \pi^2 
\frac{( s-2m^2 - \sqrt{E^4 - t s} )^4}{t \sqrt{E^4 - t s}}
\eea
and
\beq
\textrm{LS}^{(s ++,++)}_{\Box h} = \textrm{LS}^{(s --,--)}_{\Box h}
= i G^2 \pi^2 \frac{m^8}{t \sqrt{E^4 - t s}} .
\eeq
Now let us evaluate the one-loop leading singularity with only gravitational Merlins in the loop. By using the same technique as above, we find that
\beq
\textrm{LS}^{(s)}_{M} =  \frac{x_1}{4 m^2 (1-x_1^2)} 
\frac{1}{(2\pi i)} \oint_{\Gamma} 
\frac{dz} {z}
M_{3}({\bf 1}^{\prime}, \boldsymbol\ell, - \boldsymbol\ell_1) 
M_{3}(-\boldsymbol\ell, {\bf 1}, -\boldsymbol\ell_3) 
M_{4}({\bf 2},\boldsymbol\ell_3, \boldsymbol\ell_1, {\bf 2}^{\prime})
\eeq
with
\beq
A = - B = \frac{ ( 2 - M^2/m^2 ) x_1}{ z(1-x_1) } .
\eeq
Using the double-copy prescription enables one to derive simple expressions for the product of the amplitudes that enter in the above formula. We stress again that we are taking into account all the degrees of freedom comprised by the double copy, that is, gravitons, Merlins, dilatons and Kalb-Ramonds. In order to consider only gravitons and gravitational Merlins, one must disentangle them from the other contributions -- this is achieved by using the four-dimensional physical state projectors given above. In any case, all terms above will contribute to the triangle leading singularity. By taking 
$\Gamma = S^{1}_{\infty}$, we find that
\bea
\textrm{LS}^{(s)}_{\triangle M} &=&  
\frac{i \kappa^4}{4 m} \frac{1}{\sqrt{-t}} \frac{1}{\sqrt{4-t/m^2}} 
\frac{1}{(2\pi i)} \oint_{S^{1}_{\infty}} \frac{dy}{ y }
 \left\{ - \Biggl[  \left( s-2m^2 -  \frac{1}{2} \bigl( S(y) + M^2 \bigr) \right)^2
 + \frac{1}{8} \left(-8 m^2+2 M^2+t\right) S(y) \Biggr]^2 \frac{1}{S(y)}
\right.
\nn\\
&-& \left. \Biggl[ \left( u-2m^2  - \frac{1}{2} \bigl( U(y) - M^2 \bigr) \right)^2 
+ \frac{1}{8} \left(-8 m^2+2 M^2+t\right) U(y)
 \Biggr]^2 \frac{1}{ U(y) }
\right.
\nn\\
&+& \left. \Biggl( \frac{1}{8} \left( -8 m^2+2 M^2+3t \right) \bigl( S(y) - U(y) \bigr)  
- \frac{t}{2} ( s - u  - M^2 ) \Biggr)^2
\left( \frac{ 1 }{t} - \frac{ 2 }{t - M^2} \right)
\right\}
\eea
where
\bea
S(y) &=& \frac{1}{y}
 \biggl( \frac{E^4 r(x_1)}{2 (1-x_1)^2}  y^2
+  \frac{s \left(2 M^2-t\right)}{s+u}  y
+ 2 r(x_1) x_1\left( 1+\frac{(M^2-4 m^2) M^2 }{m^2 t} \right) \biggr)
\nn\\
U(y) &=& \frac{1}{y}
\biggl( - \frac{E^4  r(x_1)}{2 (1-x_1)^2} y^2
 + \frac{u \left(2 M^2-t\right)}{s+u}  y
- 2 r(x_1) x_1\left( 1+\frac{(M^2-4 m^2) M^2 }{m^2 t} \right)  \biggr)
\eea
and we have used the change of variables $z = r(x_1) (q \cdot p_2) y = - r(x_1) (q \cdot p_2^{\prime}) y$. The Mandelstam variables are now given by Eq.~(\ref{mandelstam}). The contour integral is not terribly difficult to solve; however, the explicit result is not particularly enlightening so we choose to leave it in such a compact form. Nevertheless, we can check from the above expression the emergence of the standard non-analytic terms $1/\sqrt{-t}$ and $1/\sqrt{4-t/m^2}$ for triangle topologies, together with the pole at $t=M^2$ which is a typical feature due to the presence of Merlin modes.

To compute the box topology, we proceed as before; that is, we consider a contour encircling one of the poles associated with the roots of the quadratic factors in the denominators. We obtain
\beq
\textrm{LS}^{(s)}_{\Box M} = - i \left( \frac{\kappa}{2} \right)^4
\frac{1}{4} \frac{1}{\sqrt{-t}} 
\left[ 
\frac{ \left(4 m^2-M^2-2 s\right)^4}{\sqrt{ \left(E^4-s t\right) \left(u-(t+u) \left(1-\frac{2 M^2}{t+u}\right)^2\right) } }
+ \frac{  \left(4 m^2-M^2-2 u\right)^4}{\sqrt{ \left(E^4-u t\right) \left(s-(t+s) \left(1-\frac{2 M^2}{t+s}\right)^2\right) } }
\right] .
\eeq
As in the Yang-Mills case, Merlin modes change dramatically the analytic structure of the amplitude -- in the present case, this non-trivial change means that, instead of the $1/t$ pole for the box topology found in the case of gravitons running in the loop, we find the non-analytic term $\sqrt{-t}$.

Now let us calculate the crossed terms. Using the same technique as above, we find that
\beq
\textrm{LS}^{(s)}_{hM} =  - \frac{x_1}{4 m^2 \left(1-x_1^2\right) } 
\sum_{h}
\frac{1}{(2\pi i)} \oint_{\Gamma} 
\frac{dz}{ z }
M_{3}({\bf 1}^{\prime}, \boldsymbol\ell, - \ell_1^{h}) 
M_{3}(-\boldsymbol\ell, {\bf 1}, -\boldsymbol\ell_3) 
M_{4}({\bf 2},\boldsymbol\ell_3, \ell_1^{-h}, {\bf 2}^{\prime})
\eeq
with the associated $A$ and $B$ given by Eq.~(\ref{96}) and
\beq
\textrm{LS}^{(s)}_{Mh} = \frac{x_1}{4 m^2 \left(1-x_1^2\right)} 
\sum_{h}
\frac{1}{(2\pi i)} \oint_{\Gamma} 
\frac{dz}{z}
M_{3}({\bf 1}^{\prime}, \boldsymbol\ell, - \boldsymbol\ell_1) 
M_{3}(-\boldsymbol\ell, {\bf 1}, -\ell_3^{h}) 
M_{4}({\bf 2},\ell_3^{-h}, \boldsymbol\ell_1, {\bf 2}^{\prime})
\eeq
and now $A$ and $B$ are given by Eq.~(\ref{98}). In turn, the amplitudes listed above allows us to calculate the products of the amplitudes that appear in both expressions. As above we take $\Gamma = S^{1}_{\infty}$ for the calculation of the triangle topology. Performing the same change of variables as above, we find that
\bea
\textrm{LS}^{(s)}_{hM} &=&  16 i \left( \frac{\kappa}{2} \right)^4
\frac{1}{4 m} \frac{1}{\sqrt{-t}} \frac{1}{\sqrt{4-t/m^2}} 
\frac{1}{(2\pi i)} \oint_{S^{1}_{\infty}} 
\frac{dy}{ y }
\Biggl\{
  \biggl[ \left( s-2m^2 - \frac{1}{2} {\cal S}(y) \right) ( s-2m^2 )
+  \frac{  \left(t-4 m^2+M^2\right) }{4} {\cal S}(y)
\biggr]^2  \frac{1}{ {\cal S}(y)}
\nn\\
&+& 
\biggl[ \left( u-2m^2  -  \frac{1}{2} {\cal U}(y) \right) ( u-2m^2 )
+  \frac{ \left(t-4 m^2+M^2\right)}{4}  {\cal U}(y)
\biggr]^2  \frac{1}{ {\cal U}(y) }
\nn\\
&+&   
\biggl[ \frac{  \left(t-4 m^2+M^2\right) }{4} \Bigl( {\cal S}(y) - {\cal U}(y) \Bigr) 
+ \frac{t}{2} {\cal S}(y)  +  \frac{1}{4} t (-3 s+t+3 u) \biggr]^2 \frac{1}{t-M^2}
\Biggr\}
\eea
and
\bea
\hspace{-20mm}
\textrm{LS}^{(s)}_{Mh} &=& - 16 i \left( \frac{\kappa}{2} \right)^4
\frac{1}{4 m} \frac{1}{\sqrt{-t}} \frac{1}{\sqrt{4-t/m^2}} 
\frac{1}{(2\pi i)} \oint_{S^{1}_{\infty}} 
\frac{dy}{y}
\nn\\
&\times& \Biggl\{
\biggl[ \left( u-2m^2  - \frac{1}{2} {\cal U}(y) \right) \left( u-2m^2\right)
+ \frac{\left( t - 4m^2 \right)}{4} {\cal U}(y) \biggr]^2
\frac{1}{{\cal U}(y)}
\nn\\
&& + \biggl[ \left( s-2m^2 - \frac{1}{2} {\cal S}(y)  \right) \left( s-2m^2 \right)
+ \frac{ (t-4m^2)}{4} {\cal S}(y) \biggr]^2
\frac{1}{ {\cal S}(y) }
\nn\\
&& +  \biggl[  \frac{t}{2} \Bigl( {\cal S}(y) - {\cal U}(y) \Bigr) - 2 m^2 {\cal S}(y) 
-\frac{1}{4} t (4 s+t-2 u) - \frac{M^2}{4} \left( t - 4 m^2  \right)  \biggr]^2
\frac{1}{t-M^2} 
\Biggr\}
\eea
where
\bea
{\cal S}(y) &=&  \frac{1}{y} 
 \left(  - \frac{(t-M^2)}{s+u}  s y
+ \frac{E^4}{2 (1-x_1)^2} r(x_1)  y^2 
+ 2 r(x_1) x_1 \frac{ \left(t-M^2\right)^2  }{t^2}  \right)
\nn\\
{\cal U}(y) &=& \frac{1}{y}  
 \left( - \frac{(t-M^2)}{s+u } u  y 
- \frac{E^4}{2(1-x_1)^2} r(x_1)  y^2
- 2 r(x_1) x_1 \frac{ \left(t-M^2\right)^2  }{t^2}  \right) .
\eea
Again the emergence of the non-standard pole at $t=M^2$ is manifest.

The box topology for the crossed terms can be calculated in much the same way as before; we obtain that
\beq
\textrm{LS}^{(s)}_{\Box hM} =   \frac{i \kappa^4}{4} \frac{\sqrt{(4m^2-t)^2}}{\sqrt{4m^2-t}} 
\Biggl[  \frac{\left(s-2 m^2\right)^4}{\sqrt{m^2 s \left(M^2-t\right)^2 (-s t-4 u)} }
+ \frac{\left(u-2 m^2\right)^4}{\sqrt{m^2 u \left(M^2-t\right)^2 (-u t-4 s)} }
\Biggr]
\eeq
and
\beq
\textrm{LS}^{(s)}_{\Box Mh} = - \frac{i \kappa^4}{4} \frac{\sqrt{(4m^2-t)^2}}{\sqrt{4m^2-t}} 
\Biggl[  \frac{\left(u-2 m^2\right)^4}{\sqrt{m^2 u \left(M^2-t\right)^2 (-4 s-t u)} }
+ \frac{\left(s-2 m^2\right)^4}{\sqrt{m^2 s \left(M^2-t\right)^2 (-4 u-t s)} }
\Biggr] .
\eeq
Similar to the gauge-theory case, the sum of both contributions also vanishes.

\section{Summary}

Unitarity cut enables one to establish a relationship between the pole structure of the integrand and the branch-cut structure of the loop integral. This allows one to retrieve the loop integrand by analyzing different sets of unitarity cuts. It turns out that the problem of finding functions which can reproduce all such discontinuities can be a daunting endeavor. We have studied here one special class of such singularities, the leading singularities, which are the ones with the highest co-dimension. We have investigated leading singularities in a higher-derivative Yang-Mills theory and quadratic gravity. As shown in previous sections, this technique can still be a powerful one to address the reconstruction of loop amplitudes for such theories.  Indeed, the applicability of unitarity-based methods to amplitudes involving unstable particles can be envisaged as a consequence of unitarity. As proved in Refs.~\cite{Veltman:63,DM:19}, theories with unstable particles are unitary. Therefore, unitarity methods must be trustworthy in such cases. This is extensively discussed in Ref.~\cite{Menezes:2021tsj}. In turn, when computing leading singularities, the associated contours  must enclose one-particle poles so that the result of the contour integration will represent an on-shell particle carrying positive energy. In the case of unstable particles, when one is off resonance, then one is also away from the pole of the propagator, and one obtains a vanishing residue. Hence a finite result is only warranted when one is close to the resonance region.

The present study is a natural continuation of the investigation initiated in the works~\cite{Menezes:2021tsj,Menezes:2021dyp} where analytic properties of scattering amplitudes of higher-derivative field theories were approached. This is an important point as it is well known that resummed Merlin propagators do not satisfy standard analyticity properties since the poles are on the physical sheet, which leads to a unorthodox analytic structure for the associated amplitudes. Indeed, as shown by the calculations above, the presence of Merlin modes alter the form of the box leading singularities in a non-trivial way. Our results suggest that a better understanding of the analytic features of the amplitudes calculated in such theories is a pressing issue. The use of tools coming from the modern on-shell amplitudes program seems to be a promising pathway to be explored. 

It is unclear whether traditional field-theory techniques agree in their results when applied to higher derivative theories. For instance, the equivalence of Euclidean and Lorentzian formulations has been grounded in standard theories and it is not clear how such correspondence would be possible
for higher derivative theories, particularly bearing mind the unconventional analytic features such theories possess~\cite{Donoghue:2021cza}. Perhaps the traditional representation of loop amplitudes in terms of Feynman diagrams might not be useful here. It is known that in many situations the Lee-Wick contour is mandatory; however, one is unsure on the consistency of this prescription to higher-orders in perturbation theory~\cite{Aglietti:2016pwz,Anselmi:2017yux,Anselmi:2017lia,Anselmi:2018kgz,Anselmi:2018tmf,Anselmi:2021hab}. In turn, it is also unclear how to include the CLOP prescription into a Lagrangian and ambiguities are expected to arise at higher orders~\cite{Boulware:1983vw}. Standard Feynman diagrams were originally conceived for theories with scattering amplitudes which are analytic in the first Riemann sheet, and models enjoying one unambiguous arrow of causality. This is not the case here. An alternative treatment could probe other options; perhaps recent ideas comprising representations of amplitudes in terms of contour integrals in Grassmannian spaces or geometrizations such as amplituhedrons could be helpful here~\cite{Elvang:15,Arkani-Hamed:2009ljj,Hodges:2009hk,Arkani-Hamed:2010wgm,Arkani-Hamed:2013jha,Arkani-Hamed:2017mur}. On the other hand, as discussed, gravitational scattering amplitudes can be written as an appropriate factorization in terms of gauge-theory amplitudes. The conjecture is that the double copy might be a general imprint of all gravitational theories. The current study as well as the investigation put forward in Ref.~\cite{Menezes:2021dyp} were restricted to examine in detail $4$-point amplitudes in quadratic gravity; it would be interesting to see more evidence in favor of a double-copy structure with five or higher-point amplitudes. These are certainly explorations worthy of further consideration within the program of quadratic gravity as a potential UV completion for quantum gravity, and we hope to return to such considerations in the near future.

\section*{Acknowledgements} 

This work has been partially supported by Conselho Nacional de Desenvolvimento Cient\'ifico e Tecnol\'ogico - CNPq under grant 317548/2021-2 and Funda\c{c}\~ao Carlos Chagas Filho de Amparo \`a Pesquisa do Estado do Rio de Janeiro - FAPERJ under grants E- 26/202.725/2018 and E-26/201.142/2022.

\end{document}